\documentclass[sigconf]{acmart}
\AtBeginDocument{%
  }

\copyrightyear{2025} 
\acmYear{2025} 
\acmConference[CHI '25]{CHI Conference on Human Factors in Computing Systems}{April 26-May 1, 2025}{Yokohama, Japan}
\acmBooktitle{CHI Conference on Human Factors in Computing Systems (CHI '25), April 26-May 1, 2025, Yokohama, Japan}\acmDOI{10.1145/3706598.3713381}
\acmISBN{979-8-4007-1394-1/25/04}

\usepackage{booktabs}
\usepackage{multirow}
\usepackage{makecell}
\usepackage{cleveref} 
\usepackage{listings}
\usepackage{enumitem}
\newcommand{\xiyuan}{\textcolor{black}}
\newcommand{\change}{\textcolor{black}}

\makeatletter
\def\adl@drawiv#1#2#3{%
        \hskip.1\tabcolsep
        \xleaders#3{#2.3\@tempdimb #1{1}#2.3\@tempdimb}%
                #2\z@ plus1fil minus1fil\relax
        \hskip.1\tabcolsep}
\newcommand{\cdashlinelr}[1]{%
  \noalign{\vskip\aboverulesep
           \global\let\@dashdrawstore\adl@draw
           \global\let\adl@draw\adl@drawiv}
  \cdashline{#1}
  \noalign{\global\let\adl@draw\@dashdrawstore
           \vskip\belowrulesep}}
\makeatother

\newcommand{\Classification}[1]{\textbf{\textcolor[RGB]{204, 120, 47}{#1}}}
\newcommand{\Understanding}[1]{\textbf{\textcolor[RGB]{84, 129, 100}{#1}}}
\newcommand{\Interpretation}[1]{\textbf{\textcolor[RGB]{138, 103, 171}{#1}}}

\begin{document}

\title{ClueCart: Supporting Game Story Interpretation and Narrative Inference from Fragmented Clues}

\author{Xiyuan Wang}
\orcid{0009-0008-1839-2010}
\affiliation{%
    \institution{School of Information Science and Technology, ShanghaiTech University}
  \city{Shanghai}
  \country{China}
}
\email{wangxy7@shanghaitech.edu.cn}

\author{Yi-Fan Cao}
\orcid{0000-0002-5892-5052}
\affiliation{%
  \institution{Academy of Interdisciplinary Studies, The Hong Kong University of Science and Technology}
  \city{Hong Kong}
  \country{China}}
\email{ycaoaw@connect.ust.hk}

\author{Junjie Xiong}
\orcid{0009-0004-6005-7508}
\affiliation{%
    \institution{School of Information Science and Technology, ShanghaiTech University}
  \city{Shanghai}
  \country{China}
}

\author{Sizhe Chen}
\orcid{0009-0004-5612-9935}
\affiliation{%
  \institution{Physics, Mathematics and Environment, Chalmers University of Technology}
  \city{Gothenburg}
  \country{Sweden}
}
\email{sizhec@chalmers.se}

\author{Wenxuan Li}
\orcid{0009-0008-5210-9995}
\affiliation{%
  \institution{University of Southern California}
  \city{Los Angeles}
  \state{California}
  \country{USA}}

\author{Junjie Zhang}
\orcid{0000-0002-3155-5805}
\affiliation{%
  \institution{Information Hub, The Hong Kong University of Science and Technology (Guangzhou)}
  \city{Guangzhou}
  \country{China}}
\email{jakezhang@hkust-gz.edu.cn}

\author{Quan Li}
\authornote{Corresponding Author.}
\orcid{0000-0003-2249-0728}
\affiliation{%
  \institution{School of Information Science and Technology, ShanghaiTech University}
  \city{Shanghai}
  \state{}
  \country{China}
  }
\email{liquan@shanghaitech.edu.cn}

\renewcommand{\shortauthors}{Wang et al.}

\begin{abstract}
Indexical storytelling is gaining popularity in video games, where the narrative unfolds through fragmented clues. This approach fosters player-generated content and discussion, as story interpreters piece together the overarching narrative from these scattered elements. However, the fragmented and non-linear nature of the clues makes systematic categorization and interpretation challenging, potentially hindering efficient story reconstruction and creative engagement. To address these challenges, we first proposed a hierarchical taxonomy to categorize narrative clues, informed by a formative study. Using this taxonomy, we designed \textit{ClueCart}, a creativity support tool aimed at enhancing creators' ability to organize story clues and facilitate intricate story interpretation. We evaluated \textit{ClueCart} through a between-subjects study (N=$40$), using \textit{Miro} as a baseline. The results showed that \textit{ClueCart} significantly improved creators' efficiency in organizing and retrieving clues, thereby better supporting their creative processes. Additionally, we offer design insights for future studies focused on player-centric narrative analysis.
\end{abstract}


% \ccsdesc[500]{}
% \ccsdesc[300]{}



\maketitle

\section{Introduction}

\par
\begin{figure*}[h]
  \centering
  \includegraphics[width=\linewidth]
  {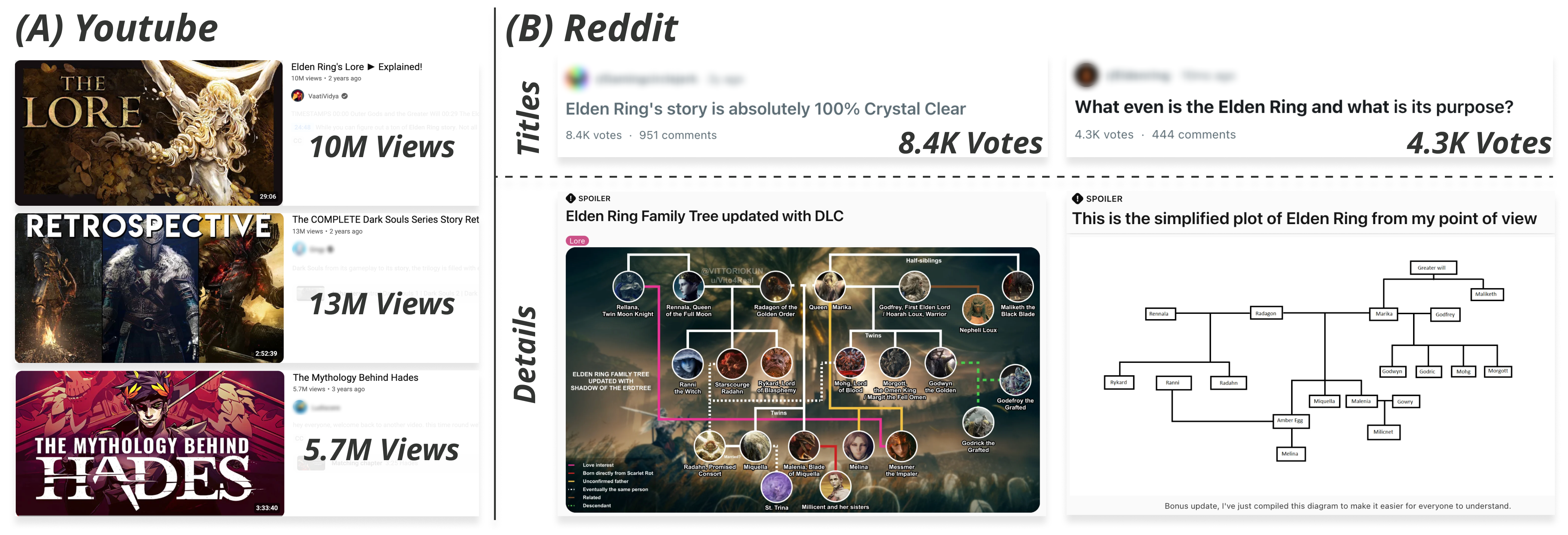}
  \caption[]{Examples of game story interpretations on (A) Youtube and (B) Reddit.}
  \label{fig:video_community}
\end{figure*}

 In video games, the story often serves as a significant selling point, providing players with an immersive experience within the game world~\cite{bormann2015immersed,somerdin2016game,christou2014interplay}. It drives player-generated content and discussions, with interpretations of game stories~\cite{cardona2020gfi} becoming increasingly popular through videos and forum discussions~\cite{anderson2018extraludic,larsen2023communal,caracciolo2024soulsring}. \xiyuan{On platforms like \textit{YouTube}\footnote{\url{https://www.youtube.com/}} and \textit{Reddit}\footnote{\url{https://www.reddit.com/}}, these \textbf{players}---who also act as content \textbf{creators} by producing and sharing their interpretations through video and forum posts---contribute to a vibrant discourse. \change{Here, content creators refer specifically to those analyzing game narratives, rather than those of user-generated content within games or game worlds.} Their content, which garners millions of views (\cref{fig:video_community}(A)) and votes (\cref{fig:video_community}(B)), reflects the depth of player engagement with game stories.} 

% Content on platforms like YouTube\footnote{\url{https://www.youtube.com/}} has garnered millions of views (\cref{fig:video_community}(A)), and Reddit\footnote{\url{https://www.reddit.com/}} posts receive numerous votes (\cref{fig:video_community}(B)), reflecting players' immense enthusiasm for game story interpretations, enhancing their immersion and engagement.

\par Popular games often employ indexical storytelling~\cite{fernandez2011game} to deliver their narratives. This technique uses \textit{indices}—elements within the game world such as items and environments that physically point to events or happenings—as clues to invite players to reconstruct the story. In \textit{Elden Ring}~\cite{gameEldenRing}, for instance, players might find cryptic messages left by other characters or environment hints about the lore. Without explicit answers from designers, this approach gives players the freedom to develop their own understanding of game events, leading to varying interpretations~\cite{fernandez2011game,livingstone2016archaeological,rassalle2021archaeogaming, andriano2024enjoying}. The enigmatic nature of these stories and the diversity of viewpoints fuel players' enthusiasm for discussing game narratives. This, in turn, motivates \xiyuan{players, acting as creators,} to publish their interpretations, contributing to a rich and diverse content ecosystem.

\par However, indexical storytelling presents significant challenges for creators of interpretations. The volume and diversity of narrative clues, ranging from environments to item descriptions, make it difficult to \textit{\textbf{organize}} them effectively. The sheer number of these clues can be overwhelming. Additionally, its fragmented structure requires substantial effort to \textit{\textbf{interpret}} the narrative clues holistically. Since the story elements are scattered throughout the game world, creators must invest considerable time and analytical skill to piece together a coherent narrative.

\par Addressing the challenge of \textit{\textbf{organization}}, current game narrative research primarily classifies narrative elements from a game designer's perspective~\cite{domsch2013storyplaying,wei2010time,wei2011analyzing,cavazza2017introduction,brandse2022conveyance,silva2019narrative,wei2010embedded}. Designers focus on creating coherent and immersive narratives, emphasizing structure and progression. In contrast, interpretation creators aim to engage their audience by presenting the story from various perspectives, exploring narrative elements non-linearly and interpretively. The designer-centric classifications do not support this exploratory and interpretive nature, resulting in heavy manual labor for creators when must organize and connect these clues to form a comprehensive understanding. 
\par Concerning the challenge of \textit{\textbf{interpretation}}, current Human-Computer Interaction (HCI) research focuses on the effectiveness of creativity support tools (CSTs) in underpinning user creative practices~\cite{shneiderman2007creativity,hwang2022too,frich2018hci}. These tools, widely used in story creation processes~\cite{lu2011shadowstory,cao2010telling,maiden2018making}, emphasize story analysis and creative thinking. They primarily utilize visualization methods such as storyline~\cite{hulstein2022geo,tang2020plotthread,tang2018istoryline}, timeline~\cite{offenwanger2023timesplines,brehmer2016timelines,thiry2013authoring,hulstein2022geo}, and storytelling~\cite{dusi2016playthecityre,morth2022scrollyvis,tong2018storytelling}. However, these tools are unsuitable for non-linear interpretation processes and are designed more for creating new stories rather than interpreting existing narrative clues. Concept maps are another type of CST used to stimulate idea generation and aid creativity~\cite{meng2016hynote,villalon2008concept,shneiderman2007creativity,simper2016effects,wang2018effects}. They are effective for note-taking and summarizing, helping users glean key concepts, relationships, and hierarchies from documents and source materials~\cite{chang2002effect}. However, they are not equipped for the dynamic and interpretive processes required in game narrative construction. These gaps underscore the need for specialized tools for game story interpretation, supporting story analysis, creative thinking, and the effective organization and interpretation of narrative clues.

\par The gaps between existing classifications and tools underscores the urgent need for research that addresses the complexities of player-driven interpretation. There is a necessity for tools and methodologies that assist creators in effectively organizing, understanding, and interpreting narrative clues. Given these challenges in the current landscape of game story interpretation, we propose investigating the following research questions (\textbf{RQs}) to bridge these gaps:
\begin{itemize}
\item \textbf{RQ1:} What taxonomy of narrative clues is most applicable for interpretation creators? 
\item \textbf{RQ2:} How do creators interact with the proposed tool?
\item \textbf{RQ3:} To what extent does the proposed tool improve the efficiency of creators' interpretation processes?
\end{itemize}

\par To address these questions, we began by exploring how creators classify and organize game clues through a combination of methods: a literature review, an application survey, and a co-design workshop involving $14$ participants. From this, we identified their common classification habits, the challenges they encounter, and distilled the design requirements for our tool. 
Building on these insights, we derived a clue taxonomy and developed an in-game clue collection method based on it. We then created \textit{ClueCart}, a creativity support tool that helps creators organize, understand, annotate, and explore relationships between various clues, thereby supporting the entire story interpretation workflow.

\par A between-subjects study with $40$ creators \xiyuan{compared \textit{ClueCart} with the baseline, \textit{Miro}, a widespread tool in creative fields, especially for story mapping~\cite{berendsen2018digital,groshans2019digital}. For quantitative evaluation, we assessed multiple dimensions: perception, including usability and creativity support, specific features of \textit{ClueCart}, and story interpretation outcomes, involving completion time and quality. We also conducted qualitative interviews to gain deeper insights into users' experiences.} The results confirmed the validity of the taxonomy and demonstrated improvements in both the efficiency and quality of story interpretation. In summary, our contributions are: 
\begin{itemize}
\item We explore narrative clues taxonomy through a literature review, an application survey, and a co-design workshop.
\item We develop an open-source creativity support tool, \textit{ClueCart}\footnote{\url{https://cluecart.github.io/ClueCart/}}, along with a clue collection method, enabling creators to organize, understand, and interpret game stories based on fragmented clues.
\item We evaluate \textit{ClueCart} through a between-subjects study (N=$40$), \xiyuan{comparing it a baseline, \textit{Miro}\footnote{\url{https://miro.com/app/dashboard/}},} and offer design implications for future studies on player-centric narrative analysis.
\end{itemize}

\section{Background and Related Work}
\subsection{Background}
\subsubsection{Game Story Interpretations}
\par Game story interpretation via videos and forums has surged in popularity. Platforms like \textit{YouTube} host numerous videos analyzing intricate game stories, attracting millions of views (\cref{fig:video_community}(A)). Games like \textit{Elden Ring}, \textit{Dark Souls}~\cite{gameDarkSouls}
%\footnote{A dark fantasy action role-playing game series developed by FromSoftware and published by Bandai Namco Entertainment: \url{https://www.darksouls.jp/}} 
and \textit{Hades}~\cite{gameHades}
%\footnote{A roguelike action role-playing game developed and published by Supergiant Games: \url{https://store.privatedivision.com/game/buy-hades}} 
have detailed analysis videos demonstrating players' deep interest in game stories. Similarly, \textit{Reddit threads} allow players to exchange insights, enhancing collective understanding (\cref{fig:video_community}(B)).

\par Videos play a crucial role in game story interpretation, ranging from simple walkthroughs to in-depth analyses. These videos explore lore, character motivations, and mysterious plots, making complex narratives accessible to a broader audience through engaging visual and auditory content. Community forums and social media platforms further facilitate the collective interpretation of game stories. \textit{Subreddits} dedicated to specific games serve as hubs where players dissect narrative elements and propose interpretations, building on each other's ideas to form comprehensive understandings. Game story interpretation through videos and community discussions significantly enhance player engagement and immersion~\cite{bormann2015immersed,somerdin2016game,christou2014interplay}. 

% Improving creators' interpretation efficiency can better support this vibrant aspect of gaming culture.

\par \xiyuan{Digital whiteboards, such as \textit{Miro}, are widely employed in creative fields to support collaborative brainstorming and story mapping~\cite{berendsen2018digital,groshans2019digital}, where visualizing complex narratives is essential. Creators often rely on these tools to manually arrange screenshots, notes, and diagrams that represent fragmented narratives, creating ad-hoc structures to analyze and communicate their interpretations. Research has demonstrated the effectiveness of such tools in organizing intricate ideas and enhancing team collaboration~\cite{radics2021methodological,klein2023reimagining}, making them ideal for creators to better interpret and communicate multifaceted narratives in video games. However, as the complexity and number of story fragments in modern games grow, these tools often struggle to manage the increasing intricacies of game story analysis. Our work advances such tools by proposing and integrating a taxonomy for the automatic organization and retrieval of game story clues.}

\subsubsection{Indexical Storytelling}
\par Besides the longstanding academic debate between Ludology and Narratology~\cite{koenitz2024narrative,veloya2023analysis,murray2005last,frasca2013simulation,frasca1999ludology}, which explores whether video games should be understood primarily through their mechanics or as narrative mediums, recent research has focused on interactive forms of narrative~\cite{ferri2007making,ferri2007narrating,ferri2013satire,ferri2015narrative,mateas2001preliminary,mateas2005structuring}. One notable form is indexical storytelling, introduced by Fernandez et al.~\cite{fernandez2011game}, who applied a semiotic perspective based on Charles Peirce's philosophy of language~\cite{lycan2018philosophy}. Building on the approaches of Nitsche~\cite{nitsche2008video} and presentations by Rouse~\cite{rouse2010environmental}, Smith and Worch~\cite{smith2010happened} at the Game Developers Conference in San Francisco, Fernandez et al. conceptualized ``indexical storytelling'' through meaningful objects, encounters, and traces left by other players.

%Indexical storytelling constructs narratives through indices, encouraging players to reconstruct game events from clues. 

\par \xiyuan{In games using indexical storytelling, players are not merely passive recipients of a narrative. They are invited to interpret clues---whether physical objects, environmental details, or traces of past player actions---to piece together a cohesive story. The inherently mysterious nature of these narratives creates substantial room for interpretation, motivating players, as creators, to share their unique perspectives and analyses within the gaming community~\cite{larsen2023communal,anderson2018extraludic}. Game examples} include detective works in \textit{Phoenix Wright: Ace Attorney}~\cite{gamePhoenixWright},
% \footnote{A visual novel adventure game developed and published by Capcom: \url{https://www.ace-attorney.com/}}
remains interpretation in \textit{BioShock}~\cite{gameBioShock},
% \footnote{A first-person shooter game developed by 2K Boston and 2K Australia, and published by 2K: \url{https://2k.com/en-US/game/bioshock-the-collection/}}
and signage and tutorials in \textit{Super Mario 64}~\cite{gameSuperMario64}.
% \footnote{A platform game developed and published by Nintendo: \url{https://www.nintendo.com/en-gb/Games/Nintendo-64/Super-Mario-64-269745.html}}
Additionally, it reconstructs player history through traces left by players, as seen in \textit{Demon's Souls}~\cite{gameDemonsSouls}.
% \footnote{An action role-playing game developed by FromSoftware: \url{https://www.fromsoftware.jp/ww/detail.html?csm=070}}. 
Subsequent works have further explored the archaeological aspects of indexical storytelling~\cite{livingstone2016archaeological,smith2023darned} and expanded its scope. Warpefelt et al.~\cite{warpefelt2016non} argued that NPCs (non-playable characters) fit within this framework, with their attributes and visual presentations enriching the narrative experience for players. Our work builds on the concept of indexical storytelling, examining how story interpreters understand and classify this form of storytelling, thereby enhancing the narrative experience and engagement in video games.

\subsection{Game Narrative Elements Classification}
\label{sec:game_cls}
\par Research on the classifications of game narrative elements, such as \textit{cut-scenes, dialogue, text, voice-over narration, narrative artifacts}, and \textit{environment}, includes a wide range of approaches. These elements are the foundational components of a game's story, shaping how narratives are presented and experienced by players~\cite{wei2010embedded,10.1007/978-3-030-04028-4_27,silva2019narrative,brandse2022conveyance,domsch2013storyplaying,cavazza2017introduction,wei2011analyzing}. 
%Given their varied forms, these elements exhibit multi-modal properties, integrating visual, auditory, and textual data to create immersive storytelling experiences.

\par Most research classifies narrative elements from the perspective of game story designers. For instance, Wei et al.~\cite{wei2010embedded} classify elements into horizontal, vertical, and modal embeddings based on the relationship between the embedded narrative and the main narrative. Some studies borrow models and theories from other fields~\cite{10.1007/978-3-030-04028-4_27,silva2019narrative}. For example, Silva et al.~\cite{silva2019narrative} utilized the literary theory of J. Hillis Miller~\cite{miller2005j}, categorizing elements into situation, character, and form. Furthermore, some research focuses on the interactive narrative elements~\cite{brandse2022conveyance,domsch2013storyplaying}. Brandse et al.~\cite{brandse2022conveyance} categorized elements based on their interactivity and traditional aspects, while Domsch et al.~\cite{domsch2013storyplaying} classified elements into passive, actively nodal, and dynamic categories.

% \par However, existing research predominantly reflects a designer-centric perspective, focusing on structured and progressive narrative elements. This contrasts sharply with the non-linear, interpretive methods employed by creators who piece together diverse narrative fragments to engage their audiences, often leading to significant manual effort. 
\par \xiyuan{However, existing research predominantly reflects a designer-centric perspective, focusing on how designers structure and present narrative elements\change{~\cite{wei2011analyzing,domsch2013storyplaying,wei2010embedded,silva2019narrative}}. 
It neglects the way that players, acting as creators piece together diverse narrative fragments to engage their audiences, which often leads to significant manual effort.}
Our work addresses this gap by proposing a new taxonomy that emphasizes the interpretation of indexical storytelling clues, offering a novel perspective for understanding and analyzing complex narratives in games.

% \par \xiyuan{Additionally, given the varied forms of the elements, they exhibit multi-modal properties, integrating visual, auditory, and textual data to create immersive storytelling experiences. The complexity and volume of these indices highlight the need for automated classification methods to reduce the manual effort required from creators. 
% To address these needs, our work draws on multi-modal classification methods from existing literature~\cite{}.
% % 等方法定下来了补充
% For example, [xxx]. 
% Our proposed automated classification system leverages these multi-modal techniques to categorize narrative indices in a way that aligns with the interpretive practices of game story creators. By integrating these advanced methods, we aim to develop tools that support efficient narrative organization and enrich the creative and interpretive processes in game storytelling.}

\subsection{Creativity Support Tools}
\par Creativity Support Tools (CSTs) in HCI research aim to improve creative practices by providing innovative ways for generating, organizing, and refining ideas~\cite{shneiderman2007creativity,hwang2022too,frich2018hci}. These tools are widely utilized in story analysis and creative thinking, including storytelling~\cite{dusi2016playthecityre,morth2022scrollyvis,tong2018storytelling}, timeline visualization~\cite{offenwanger2023timesplines,brehmer2016timelines,thiry2013authoring,hulstein2022geo}, and storyline visualization~\cite{hulstein2022geo,tang2020plotthread,tang2018istoryline,padia2018yarn,padia2019system,di2020storyline}. They also include other tools aimed at enhancing story creation~\cite{lu2011shadowstory,cao2010telling,maiden2018making}. For instance, Tang et al.~\cite{tang2020plotthread} utilized reinforcement learning to optimize storyline layout, aiding creators in plot analysis. Similarly, \textit{TimeSplines}~\cite{offenwanger2023timesplines}, a timeline visualization tool, emphasizes reflection, enabling creators to effectively analyze temporal story flow. Tools like \textit{StoryKit}~\cite{bonsignore2013sharing} enhance creative expression by integrating text, images, and sounds into storytelling. Collectively, these tools promote a more immersive storytelling experience. Despite these advancements, existing story analysis and creation tools primarily target the development of new narratives rather than the interpretation and reconstruction of fragmented narrative clues common in complex game storytelling. They do not address the challenges creators face when piecing together diverse narrative fragments into coherent stories.

\par Concept maps are another type of tool designed to stimulate creativity~\cite{meng2016hynote,tang2021conceptguide,villalon2008concept,le2018improving,shneiderman2007creativity,simper2016effects,wang2018effects} by organizing thoughts and ideas. Le et al.~\cite{le2018improving} tested the use of concept map planning to support creativity in photo stories, hypothesizing that organizational skills taught through concept maps would improve creative outcomes. Additionally, Wang et al.~\cite{wang2018effects} integrated the multi-modal framework of learning analytics with the concept mapping approach to enhance students' vocabulary and reading abilities. While concept maps are effective for note-taking, summarizing, and organizing thoughts~\cite{chang2002effect}, they are not adequately equipped for the dynamic and interpretive processes required in game narrative construction. Existing tools focus on improving creativity and organization but do not address the specific needs of reconstructing fragmented game narratives.

\par To fill these gaps, our approach combines concept maps with clues classification and interactions derived from formative studies. This method aids creators in organizing, analyzing, and interpreting narrative clues, facilitating the dynamic synthesis of these elements into cohesive stories, and enriching the creative and interpretive processes.

\section{Methodology}
\label{sec:formative study}
\par With approval from our institution's IRB, we conducted a formative study involving $14$ \xiyuan{players who are also content creators} experienced in sharing their interpretations of game narratives with the public. This study aimed to explore how \xiyuan{players, in their role as creators} currently interpret game stories and included two main components: (1) a literature review and an application survey (\cref{fig:formative_study} (1)), and (2) a co-design workshop (\cref{fig:formative_study} (2-6)).

\subsection{Literature Review and Application Survey}
\label{sec: literature and application}
\par
\begin{figure*}[h]
  \centering
  \includegraphics[width=1\linewidth]
  {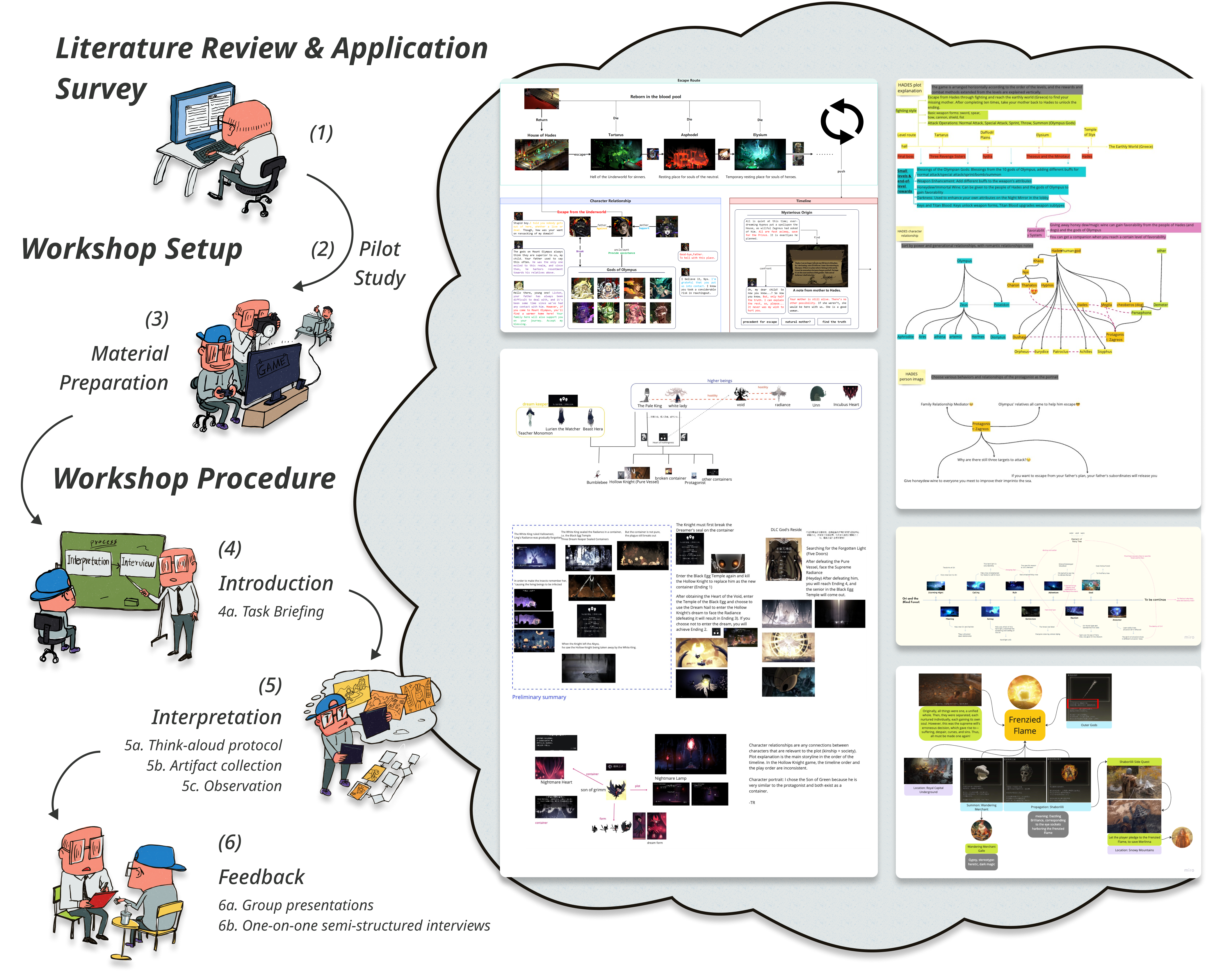}
  \caption[]{Formative Study procedure. \xiyuan{\textbf{\textit{Literature Review \& Application Survey:}} (1) Conduct a literature review and application survey to gather three narrative modalities in game story interpretation. \textit{\textbf{Workshop Setup:}} (2) Perform a pilot study to assess appropriateness. (3) Request participants to prepare game clues. \textit{\textbf{Workshop Procedure:}} (4) Introduce the research background and workshop process to participants. (5) Ask participants to complete a game story interpretation task. (6) Interview each participant to gather feedback.}}
  \label{fig:formative_study}
\end{figure*}

 Besides the literature review in \cref{sec:game_cls}, we conducted an application survey (\cref{fig:formative_study} (1)) on ``video platform commentary'' and ``forum discussions'', utilizing methodologies adapted from previous academic research~\cite{smith2023game,daneels2022digital,wulf2022content}. We gathered and categorized data from video platforms like \textit{YouTube}, as well as popular forums like \textit{Reddit} and the \textit{Steam Community}\footnote{\url{https://steamcommunity.com/}}.
\xiyuan{Based on a basic textual analysis of user comments on these popular platforms, most users express their opinions about (1) how decision-making drives the game's plot, (2) the relationships between characters, and (3) how characters express their personalities and develop through interaction.}

\par \xiyuan{We discuss the initial selection of games} from a wide range of narratives for further analysis, contrasting the narrative techniques of over $30$ well-known games on the market. This resulted in the identification of nine narrative presentation categories: \textit{Relatively Straightforward}, \textit{Visual-Driven}, \textit{Text-Driven}, \textit{Non-Linear}, \textit{Fragmented}, \textit{Lore-Rich}, \textit{Environmental Storytelling}, \textit{Branching}, and \textit{Emergent Narrative}. Two representative games from each category were chosen for further analysis, narrowing the scope to $18$ games. \xiyuan{After confirming the representativeness and diversity of the aforementioned samples, and using the classical narrative theory of ``narrative function analysis''\cite{propp1968morphology}, the stories were categorized using the fundamentals of plot, character, and setting. A classification framework that works with our research system was then developed, in which we} clearly defined which portions of the story or forum discussions would be analyzed, concentrating on particular character and narrative developments. The next phase \xiyuan{is to use the semantic analysis method Latent Dirichlet Allocation (LDA)\cite{blei2003latent} to thematically model the review data and create a coding framework based on high frequency and centrality. This approach has been validated and} summarized in three perspectives: narrative focus based on story progression, character portrayal through gameplay, and interaction patterns characterizing relationships between characters.

\begin{figure*}[h]
  \centering
  \includegraphics[width=0.935\linewidth]
  {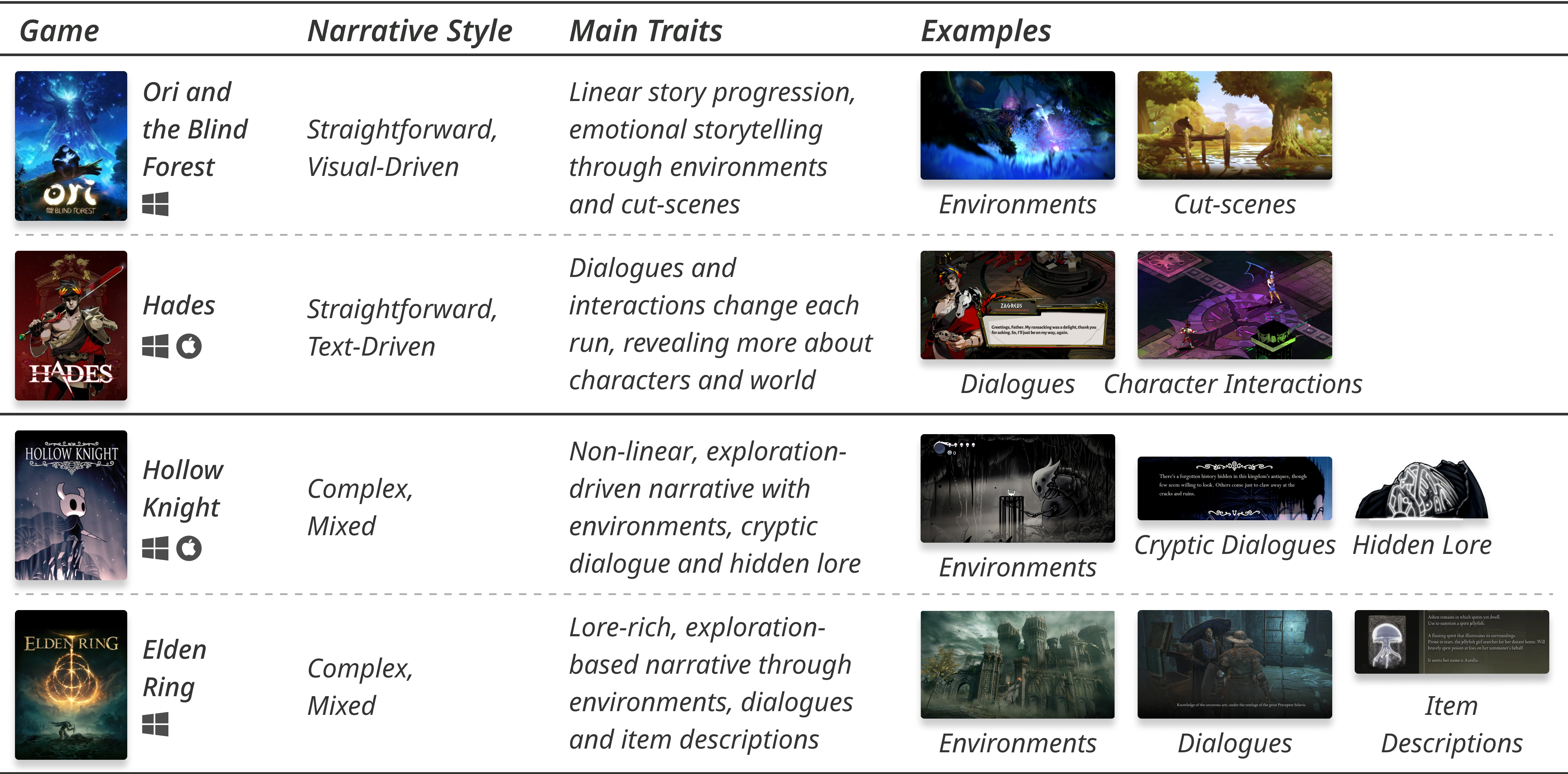}
  \caption[]{Four games adopted in the workshop using indexical storytelling. \textit{Ori and the Blind Forest} and \textit{Hades} have relatively straightforward narratives, while \textit{Hollow Knight} and \textit{Elden Ring} present more complex and mysterious narratives.}
  \label{fig:workshop_games}
\end{figure*}

\par While there are likely valid reasons for the various forms of presentation in game commentary, limited research exists on whether particular modes of narration enhance the development of game content. To address this gap, this study presents a unified classification of narrative modalities used across both research and commercial game review platforms. These modalities are categorized into three key areas:
\begin{enumerate}
    \item \textit{\textbf{Plot Analysis}}: This focuses on player choice, environmental storytelling, and non-linear narratives in video games. Narrators often explore key plot points, twists, and how gameplay choices shape the overall narrative.
    \item \textbf{\textit{Character Relationships}}: This examines the dynamic interactions between protagonists, antagonists, and other significant characters. Similar to discussions on emergent narratives in academic research~\cite{tanenbaum2009commitment}, commentators analyze player-driven connections in cooperative settings or intricate character development in story-rich games.
    \item  \textbf{\textit{Character Portrait}}: This delves into how characters evolve and interact with their environment through gameplay~\cite{smith2023game}, assessing how these portrayals contribute to the narrative depth.
\end{enumerate}
\par These categories represent widely adopted design approaches in game commentary, and we will reference them throughout this study when discussing various narrative design styles.

\begin{table*}[h]
  \caption{Detailed information of the participants in three workshops.}
  \label{tab:participants}
  \begin{tabular}{ccccccccc}
    \toprule
    \textbf{Workshop} & \textbf{ID} & \textbf{Gender} & \textbf{Age} & \textbf{Platform} & \textbf{Frequency} & \textbf{Game}  \\
    \midrule
    \multirow{3}{*}{I}  
      & P1 & Female   & 24  & Video Platforms \& Forums     & Monthly    & Hades     \\
      & P2 & Male & 22  & Video Platforms \& Forums   & Monthly        & Hollow Knight     \\
      & P3 & Male   & 20  & Video Platforms \& Forums   & Seasonally   & Ori and the Blind Forest     \\   
     % \cdashlinelr{1-7}
     \midrule
     \multirow{5}{*}{II}  
      & P4 & Male   & 22  & Video Platforms \& Forums  & Weekly  & Hades \\
    & P5 & Female   & 24  & Video Platforms \& Forums   & Monthly    & Hades     \\
      & P6 & Male   & 22  & Video Platforms \& Forums   & Monthly      & Hades     \\
      & P7 & Male   & 21  & Video Platforms   & Monthly      & Hollow Knight     \\
      & P8 & Female & 19  & Video Platforms    & Monthly      & Hollow Knight     \\
     % \cdashlinelr{1-7}
     \midrule
     \multirow{6}{*}{III}  
      & P9 & Female & 24  & Video Platforms     & Seasonally    & Ori and the Blind Forest     \\
      & P10 & Female   & 24  & Video Platforms   & Weekly   & Hades     \\
      & P11 & Male   & 24  & Video Platforms \& Forums     & Seasonally      & Hades     \\
      & P12 & Female   & 24  & Video Platforms     & Monthly    & Hades     \\
      & P13 & Male   & 29  & Video Platforms \& Forums     & Weekly      & Elden Ring     \\
      & P14 & Male   & 34  & Video Platforms \& Forums     & Weekly    & Elden Ring     \\
    \bottomrule
  \end{tabular}
\end{table*}

\subsection{Participants and Data}
\par We recruited $14$ story interpretation creators ($6$ females, $8$ males; age range $18-34$) to our workshop through online social platforms. All participants had experience publishing game story interpretations, with $5$ specializing in video content, and $9$ in both. Each participant was compensated $\$10$ per hour for their involvement.

\par Given the absence of systematic classifications, we chose not to provide participants with pre-classified game clues. Instead, we presented them with a list of potential games and asked them to select one to play prior to the workshop. The games included: \textit{Ori and the Blind Forest}, \textit{Hades}, \textit{Hollow Knight}, and \textit{Elden Ring} (\cref{fig:workshop_games}). Each game invites players to assemble its narrative, albeit in different ways. \textit{Ori and the Blind Forest}\xiyuan{~\cite{gameOri}}
% \footnote{\url{https://www.xbox.com/en-US/games/store/ori-and-the-blind-forest-definitive-edition/BW85KQB8Q31M}}
and \textit{Hades} present relatively straightforward narratives. \textit{Ori and the Blind Forest} relies on a visual-driven narrative, conveying its story primarily through stunning environments and emotional cut-scenes\xiyuan{~\cite{salonen2017practicing,mears2017design}}. \textit{Hades}, on the other hand, is text-driven, with its narrative unfolding across repeated runs, where character interactions and dialogue evolve as the player progresses\xiyuan{~\cite{shibolet2023harmony,morgan2022little}}. In contrast, \textit{Hollow Knight}\xiyuan{~\cite{gameHollowKnight}}
% \footnote{\url{https://www.hollowknight.com/}}
and \textit{Elden Ring} offer more mysterious and complex narratives. \textit{Hollow Knight} features a non-linear narrative that encourages exploration and discovery, with its story conveyed through environmental storytelling, cryptic dialogue, and hidden lore, allowing players to piece together the narrative at their own pace\xiyuan{~\cite{grunberg2024ones,raduazzo2024isolation}}. \textit{Elden Ring} employs a fragmented, lore-rich narrative style, where the story is revealed through environmental clues, item descriptions, and brief NPC dialogues\xiyuan{~\cite{spezzaferrisoulslike,larsen2023communal}}. Players are encouraged to explore and interpret the game's world, piecing together its history and lore through their discoveries.

\par Participants selected games based on their preferred platforms (e.g., \textit{Windows} and \textit{MacOS}) and personal interest in the gameplay. Their information and selected games are shown in \cref{tab:participants}. To ensure a comprehensive narrative experience, participants were required to play their chosen game for at least $4$ hours and collect a minimum of $20$ narrative-related clues during their gameplay.

\subsection{Workshop Setup}
\label{sec: workshop}
\par To ensure the workshop design was appropriate, two research team members conducted a pilot study prior to the main sessions (\cref{fig:formative_study} (2)). Given the participants' diverse geographical locations, we organized three virtual workshops using \textit{Zoom}\footnote{\url{https://zoom.us/}} for video conferencing and \textit{Miro} for collaboration.
All necessary materials, including consent forms and pre-workshop guidance, were provided five days before each session, allowing sufficient time for participants to prepare their game clues. (\cref{fig:formative_study} (3)). The guidance included the workshop background, objectives, clue collection instructions, and examples of game story interpretations. To ensure a smooth and efficient workshop experience, we sent the \textit{Zoom} and \textit{Miro} links one day before each session, giving participants time to familiarize themselves with the tools.

\subsection{Workshop Procedure}
\par The workshop was divided into three phases:  \textit{\textbf{introduction}}, \textit{\textbf{interpretation}}, and \textit{\textbf{feedback}}.

\par In the \textit{\textbf{Introduction}} phase (\cref{fig:formative_study} (4)), we briefly outlined the workshop structure, provided examples for story interpretation, and explained the tasks for the \textit{Interpretation} phase (\cref{fig:formative_study} (4a)).  Since participants had already reviewed the pre-workshop guide, this phase was brief, lasting about $10$ minutes.
During the \textit{\textbf{Interpretation}} phase (\cref{fig:formative_study} (5)), participants analyzed their narrative clues from three perspectives: \textit{plot analysis}, \textit{character relationships}, and \textit{character portrait}.
They could work digitally on \textit{Miro} or sketch on paper and upload their work. Participants followed a think-aloud protocol (\cref{fig:formative_study} (5a)) to articulate their thought processes, enabling us to capture their classification preferences, interpretive strategies, and design decisions (\cref{fig:formative_study} (5b-c)). This phase lasted about $70$ minutes. In the \textit{\textbf{Feedback}} phase (\cref{fig:formative_study} (6)), each participant presented their interpretation for $2$ minutes to the group (\cref{fig:formative_study} (6a)), followed by $15$-minute one-on-one semi-structured interviews (\cref{fig:formative_study} (6b)). The interviews were conducted after all workshops, allowing participants to observe others' approaches and methods, which enriched the discussion. We focused on three main topics: (1) clue classification strategies; (2) interpretation perspective preferences; and (3) desired tool functionalities for classification and interpretation.

\subsection{Data Analysis}
\label{sec: data_analysis}
\par We employed a systematic approach to analyze both the interview recordings and participant sketches, ensuring the thoroughness and validity of our findings. Two team members were deeply involved in the analysis process.

% \par We began by auto-transcribing all workshop and interview recordings using Otter.ai\footnote{\url{http://Otter.ai}}, followed by a thorough manual review to ensure the accuracy of the transcriptions. This step was essential to capture all nuances, including tone and pauses, accurately. Subsequently, three members of our research team deeply engaged with the data, reading through the transcripts multiple times to familiarize themselves with the content and identify initial patterns. We employed thematic analysis~\cite{} to code the transcripts, with frequent team meetings to discuss findings. To enhance the credibility of our analysis, we incorporated triangulation by comparing results across different researchers and methods. This included cross-verifying the codes and themes identified by each researcher and integrating different perspectives to ensure the robustness of our findings.

\par We conducted an inductive thematic analysis~\cite{braun2012thematic} for the interviews. All workshop interviews were recorded, yielding approximately six hours of audio data. The analysis began with automatic transcription using \textit{Otter.ai}\footnote{\url{http://Otter.ai}}, followed by a thorough manual review to ensure accuracy, including nuances such as tone and pauses. Two researchers independently conducted a preliminary coding exercise, identifying and defining codes based on the data. They then convened to discuss their individual codes and collaboratively developed a unified coding framework, which was applied extensively. The researchers independently grouped the codes into potential themes, which were further refined through discussions to establish a clear and coherent set of themes. To complement the interview data, the same process was applied to the sketches: the researchers independently coded the sketches, identifying key visual elements and patterns. After discussing and reconciling their codes, they developed a consistent set of themes for the sketch data.

% \par After establishing themes from both the interview transcripts and sketches, we conducted a cross-referencing analysis~\cite{alhojailan2012thematic}. This involved systematically comparing the themes identified in the interviews with those derived from the sketches. For each theme, we examined its visual representation in the sketches and assessed how these visual elements aligned with or expanded upon the verbal accounts provided by participants in the interviews.

\par We then conducted a cross-referencing analysis~\cite{alhojailan2012thematic} by comparing the interview themes with those derived from the sketches, examining how visual elements aligned with or expanded upon the participants' verbal accounts. Following the principle of saturation~\cite{saunders2018saturation}, the codebook was developed using data from the first two workshops. The inclusion of sketches helped validate the interview themes, and by the third workshop, no new themes emerged, indicating that saturation had been achieved.

% \par Following the principle of saturation~\cite{saunders2018saturation}, we initially developed the codebook using data from the first two workshops. The integration of sketches into the cross-referencing process also served as a means of validating the themes identified from the interviews. By the third workshop, our analysis confirmed that no new themes emerged from either the interview data or the sketches, indicating that saturation had been achieved.

\subsection{Findings}
\par Based on the data analysis, we organized our findings into two sections: \textbf{\textit{Challenges}} and \textbf{\textit{Design Requirements}}.

\begin{table*}
  \caption{Identified challenges and corresponding design requirements.}
  \label{tab:challenges}
  \centering  % Centers the table on the page
  \begin{tabular}{llr}
    \toprule
    \textbf{Categories} & \textbf{Challenges} & \textbf{DRs}\\
    \midrule
    \multirow{2}{*}{\Classification{Classification \& Organization}} 
      & \textbf{C1}. Maintain structured classification of clues ($13/14$). & \Classification{DR1}\\
      & \textbf{C2}. Organize clues efficiently ($12/14$).  & \Classification{DR1}\\
    \midrule
    \multirow{2}{*}{\Understanding{Clue Understanding}} 
      & \textbf{C3}. Ease the cog. load of comprehending clue content ($10/14$). & \Understanding{DR2}\\
      & \textbf{C4}. Understand the relationship between clues intuitively ($9/14$). & \Understanding{DR3}\\
    \midrule
    \multirow{2}{*}{\Interpretation{Story Interpretation}} 
      & \textbf{C5}. Select relevant clues ($14/14$). & \Interpretation{DR4}\\
      & \textbf{C6}. Integrate clues with external information easily ($10/14$). & \Interpretation{DR5}\\
    \bottomrule
  \end{tabular}
\end{table*}

\subsubsection{Challenges}
\par We categorized the six challenges into three distinct areas: \textit{Clue Classification and Organization}, \textit{Clue Understanding}, and \textit{Story Interpretation} in ~\cref{tab:challenges}.

\par \Classification{Classification and Organization.} Participants unanimously agreed that the classification and organization of game clues were essential first steps in interpreting game stories. However, they encounter significant challenges in this process. 
One major difficulty lies in maintaining a structured classification due to the large volume of clues and the incremental nature of their collection (\textbf{C1}). Frequent reclassification after each collection session is time-consuming and tedious, leading to a reluctance to keep things organized. As P$1$ observed, ``\textit{When there is only a small amount of clues, not sorting it in advance does not really impact how you find [things] later. But when the volume gets large, it can quickly become overwhelming to manage.}'' 

Additionally, current methods for organizing clues are inadequate, primarily relying on chronological order, which does not align with the non-linear exploration typical in games (\textbf{C2}). This misalignment hinders effective story interpretation, as chronological organization fails to reflect the logical structure needed. P$9$ shared, ``\textit{Sometimes I explore the same place more than once. Like in Ori and the Blind Forest, I passed through Hollow Grove twice---once in the main story, and later to explore [more]. If I organize everything chronologically, clues from both visits get mixed with other areas, making it hard to focus on Hollow Grove. That's why sorting clues by location is key [in cases] like this.}''

\par \Understanding{Clue Understanding.} Participants identified significant challenges in understanding the game clues they collected, which hindered their ability to efficiently interpret the stories.
\textbf{C3} highlights the difficulty of reducing the cognitive load associated with comprehending clues content. Participants often collect clues in formats like images and videos, requiring extra effort to identify content relevant to the story. This challenge was particularly evident in text-heavy games like \textit{Hades}, where the story is primarily conveyed through dialogue and text. P$5$ explained, ``\textit{Pulling out keywords from such a huge amount of text is really tough for me. It's like in another text-heavy game, Disco Elysium---there's just so much [text] that I often feel overwhelmed, and it makes it hard to figure out the main story.}''

Another challenge, \textbf{C4}, involves intuitively understanding the relationship between clues. P$13$ emphasized the cognitive burden of recalling and connecting them: ``\textit{Sometimes I know that two items or pieces of dialogue are connected, but it's hard to track that [connection] across all the content I've collected. This is especially challenging in games where the story elements aren't directly linked but are scattered around, and you have to piece them together [yourself].}'' The participant added, ``\textit{In games like Hollow Knight or Elden Ring, where the story is scattered across different clues, figuring out how everything connects often means piecing together subtle hints. It can [get] overwhelming without some kind of system to help you spot those connections.}'' This reflects a broader issue in games with indexical storytelling: creators may struggle to synthesize disconnected narrative elements without guidance.

\par \Interpretation{Story Interpretation.} 
Participants faced significant challenges in the story interpretation process, particularly in selecting relevant clues (\textbf{C5}). The large volume of clues and the long time span over which they are collected can easily result in oversights during story interpretation. As P$14$ noted, ``\textit{The scale of Elden Ring is overwhelming, and the scattered story clues are so subtle that I often miss key details when trying to piece the narrative together.}'' This challenge is particularly common in expansive, non-linear games where the narrative is fragmented and requires players to actively collect and connect dispersed clues.

Participants also emphasized the importance of integrating clues with external information (\textbf{C6}). As P$12$ noted, ``\textit{When I'm interpreting game stories, finding connections between the game's clues and real-world history, mythology, or cultural references gives me new insights and helps me think more creatively about how this outside information can enhance the story.}'' P$7$ echoed this sentiment, stating, ``\textit{I always pay close attention to place names and architectural styles. These designs aren't just randomly made up by the game designers; they often pull from real-world references. This kind of information is super helpful for understanding the story.}''

\subsubsection{Design Requirements}
\par Based on these challenges, we identified the following design requirements to assist creators in interpreting stories more effectively.

\par \textbf{\Classification{DR1} Implement adaptive and customized classification and organization.} Structured classification of clues is crucial for enhancing the efficiency of interpreting and reconstructing game narratives. Systematic categorization, such as grouping narrative elements by type or location (\textbf{C1}), allows for quick access to critical information. Additionally, flexible organization (\textbf{C2})---whether by time, theme, or relevance---supports the seamless assembly of complex storylines, reducing cognitive load and promoting a more intuitive workflow.

\par \textbf{\Understanding{DR2} Provide clue content explanations.} Understanding complex or unfamiliar clues can be a significant challenge during the interpretation process (\textbf{C3}). To address this, the tool should offer clear explanations of each clue, including its content and keywords, to convey essential information swiftly. This feature would enable more efficient comprehension and utilization of clues in story interpretation.

\par \textbf{\Understanding{DR3} Provide clue relationship explanations.} Creators often need to understand how different clues are interconnected to effectively interpret complex narratives (\textbf{C4}). The tool should offer clear explanations of the relationships between various clues, such as how a specific dialogue relates to an item or how different scenes are linked through recurring themes. By highlighting these connections, creators can more easily piece together the broader story and avoid overlooking critical narrative links.

\par \textbf{\Interpretation{DR4} Enable quick retrieval and personalized recommendation of related clues.} Swift access to relevant clues is vital for maintaining the flow of the interpretation process (\textbf{C5}). The tool should support quick retrieval by allowing users to efficiently search for and access related clues. Additionally, recommending clues based on context or user inputs can help uncover connections that might otherwise be missed, ensuring a comprehensive and cohesive interpretation.

\par \textbf{\Interpretation{DR5} Integrate external information seamlessly.} Access to relevant external information can greatly enhance creators' interpretations of game narratives (\textbf{C6}). The tool should facilitate the seamless integration of external sources, such as historical references, mythological contexts, or cultural details, directly into the interpretation process. Providing this additional context allows creators to deepen their understanding of narrative clues and develop more informed and engaging interpretations.

\begin{figure*}[h]
  \centering
  \includegraphics[width=0.85\linewidth]
  {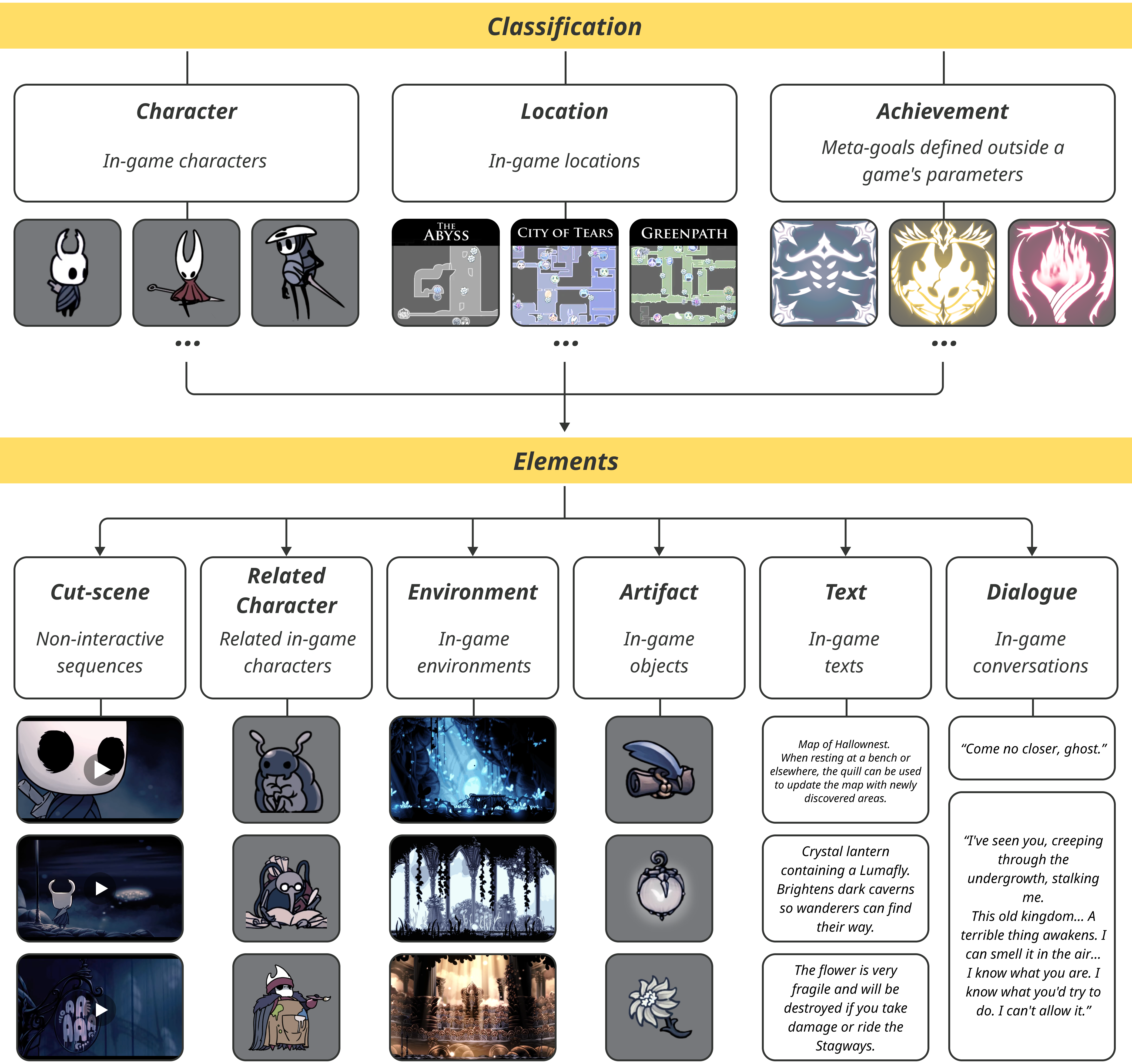}
  \caption[]{The hierarchical taxonomy, exemplified by the game \textit{Hollow Knight}, is composed of two levels. The first level, \textbf{\textit{Classification}}, categorizes clues into three groups: \textit{Character}, \textit{Location}, and \textit{Achievement}. The second level, \textbf{\textit{Elements}}, further classifies these based on in-game attributes into six types: \textit{Cut-scene}, \textit{Related Character}, \textit{Environment}, \textit{Artifact}, \textit{Text}, and \textit{Dialogue}. Each \textbf{\textit{Classification}} category can include any type of \textbf{\textit{Elements}}.}
  \label{fig:taxonomy}
\end{figure*}

\section{Taxonomy}
\par To address \textbf{RQ1}, we developed a two-level hierarchical taxonomy based on the insights collected from the workshop. \xiyuan{This taxonomy was informed by both the clues participants identified during gameplay and the insights from semi-structured interviews conducted in the co-design workshop. The clues provided concrete examples of how participants engaged with the game's narrative, while the interviews shed light on their classification strategies and reasoning. By integrating behavioral data (from the clues) and cognitive insights (from the interviews), the taxonomy offers a user-centered and flexible framework that accommodates diverse interpretations of game stories. This approach captures both what participants did and the rationale behind their interpretive choices.}

\par \xiyuan{The process, led by two team members deeply involved in the analyzing process described in \cref{sec: data_analysis}, involved two key stages. First, we inductively categorized the clues into broad thematic groups based on participants' input through open coding of both clues and interview data. Three major themes emerged from participants' approaches: \textit{\textbf{Character}}, \textit{\textbf{Location}}, and \textit{\textbf{Achievement}}. The two team members independently coded the data, then convened to discuss and develop a unified coding framework. In the second stage, we deductively identified specific narrative elements within each theme, including \textit{\textbf{Cut-scene}}, \textit{\textbf{Related Character}}, \textit{\textbf{Environment}}, \textit{\textbf{Artifact}}, \textit{\textbf{Text}}, and \textit{\textbf{Dialogue}}, based on their role in advancing the plot, conveying emotion, or providing contextual information. The interviews revealed additional nuances in how participants prioritized or combined these elements, further refining the taxonomy.}

\par \xiyuan{After categorizing and defining these elements, we organized the taxonomy into a two-level hierarchy.} The first level, \textbf{\textit{Classification}}, outlines the three common ways creators categorize game clues. The second level, \textit{\textbf{Elements}}, identifies six in-game inherent attributes of these clues. It is important to note that clues classified under any category in \textit{Classification} may encompass all types of elements outlined in \textit{Elements}.

\subsection{Classification}
\par When interpreting game stories, creators typically categorize clues retrospectively based on their gameplay experiences. We identified three key dimensions defining their categorization habits: (1) Character, (2) Location, and (3) Achievement.

\par \textbf{\textit{Classification-1: Character.}} In games, characters include both player-controlled and non-player characters (NPCs). Categorizing clues by character is crucial for retrieving content related to character portraits and relationships. This organization helps track character development, interactions, and their role in the overall narrative, ensuring consistency in portrayal and supporting a cohesive narrative analysis.

\par \textbf{\textit{Classification-2: Location.}} Locations refer to specific areas within the game world, often tied to unique storylines or events. Categorizing clues by location helps maintain geographical coherence and allows creators to analyze how the narrative unfolds across different environments, revealing the connection between setting and narrative progression.

\par \textbf{\textit{Classification-3: Achievement.}} Achievements are meta-goals linked to key in-game milestones or challenges. Categorizing clues by achievement is valuable because certain achievements relate to critical plot points. This helps map the narrative's progression and analyze how story elements are unlocked through player actions.

\subsection{Elements}
\par We identified six fundamental attributes of the assets collected by creators within the game, each contributing uniquely to the narrative structure: (1) Cut-scene, (2) Related Character, (3) Environment, (4) Artifact, (5) Text, and (6) Dialogue.

\par \textbf{\textit{Element-1: Cut-scene.}} Cut-scenes are non-interactive sequences that advance the story and provide key plot points and character development within a game. By suspending player control, they deliver the narrative with cinematic precision, ensuring intended emotional and narrative impact.

\par \textbf{\textit{Element-2: Related Character.}} Related characters are tied to specific attributes in the game and drive the narrative through interactions with the player, offering quests and revealing critical story elements. They guide or challenge the player, shaping the player's journey through dialogue and interactions.

\par \textbf{\textit{Element-3: Environment.}} Environments set the tone, mood, and atmosphere of a game, acting as narrative tools that reflect the story's progression. Environments provide visual storytelling cues, such as remnants of past events or specific objects that hint at the game's lore, influencing how players perceive and engage with the narrative and making them a vital element in narrative immersion.

\par \textbf{\textit{Element-4: Artifact.}} Artifacts are objects connected to the game's lore or history, enriching the story through exploration. These objects may be directly tied to the plot, such as an ancient weapon or a diary, or they might offer background information that enriches the world-building. They provide deeper narrative context and reward players with a richer understanding of the game world.

\par \textbf{\textit{Element-5: Text.}} In-game texts are a primary source of narrative exposition and world-building. They offer background information, character insights, and historical context, allowing players to explore the narrative at their own pace and adding depth to the story.

\par \textbf{\textit{Element-6: Dialogue.}} Dialogues are dynamic narrative components that drive character development, convey emotions, and advance the plot. Players can gain insights into characters' motivations and personalities, as well as the story's underlying conflicts and themes. Dialogue choices can influence the direction of the narrative, offering players agency in how the story unfolds and deepening their connection to the game's characters and events.

\section{ClueCart}
\par Guided by the synthesized design requirements and taxonomy, we have developed \textit{ClueCart}, an interactive creativity support tool designed to assist creators in interpreting and understanding game narratives. In the following paragraph, we will first introduce Automatic Classification section, followed by the detail functionalities of \textit{ClueCart} interface.

\subsection{Automatic Classification}
\par \textit{ClueCart}'s core functionality relies on the automatic classification of game clues according to the predefined taxonomy (\Classification{DR1}). This structured approach enhances creators' ability to analyze and interpret game narratives, revealing the intricate relationships between various game elements. To support this classification, we developed an in-game mod that automatically captures three primary categories of clues (i.e., character, location, and achievement). LLMs are then employed to further classify six additional elements, providing descriptions and extracting keywords from corresponding clues to fulfill \Understanding{DR2}.

\begin{figure*}[h]
  \centering
  \includegraphics[width=\linewidth]
  {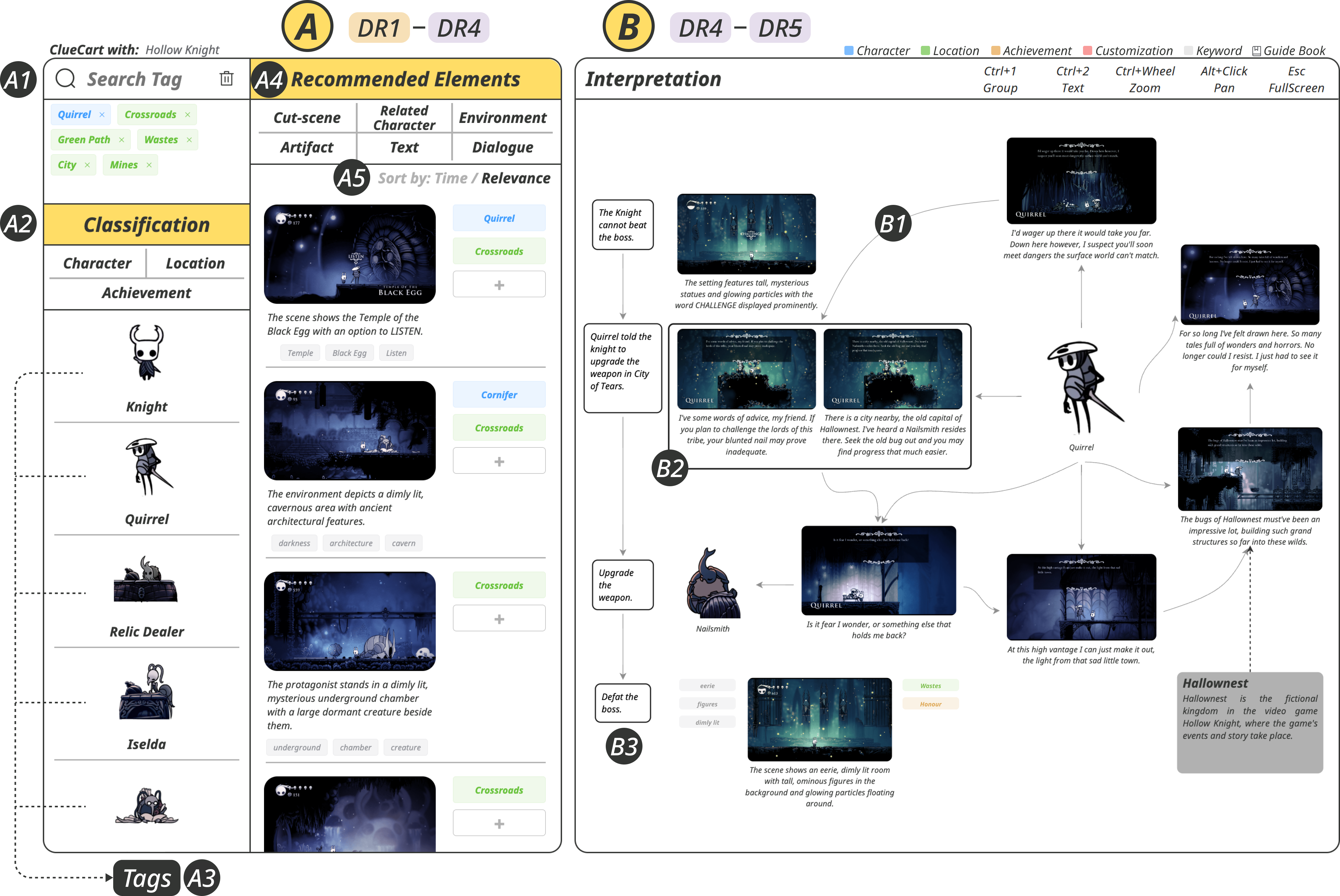}
  \caption[]{\textit{ClueCart} Interface. (A) Left panel for Clue Categorization and Retrieval. (B) Right panel for story interpretation.}
  \label{fig:cluecart}
\end{figure*}

\subsubsection{Capturing Three Classifications}
\par Video game modifications (mods)~\cite{poor2014computer} are user-generated enhancements that expand the functionality of existing games, offering players and creators new interaction possibilities. In HCI research, modding is crucial for exploring how collaborative and participatory design within gaming communities fosters innovation and enhances user experience~\cite{small2019mods,wells2018game}. We developed a custom mod using the \textit{BepInEx}\footnote{\url{https://github.com/BepInEx/BepInEx}} framework, a widely-used plugin system for \textit{Unity}-based games. This mod integrates seamlessly into the game, enabling creators to efficiently capture and classify game content. We implemented two intuitive methods based on creators' workflows: \textbf{1) \textit{Screenshots:}} Capture with a hotkey; \textbf{2) \textit{Screen Recording:}} Started and stopped with another hotkey.

\par When a capture function is activated, the mod automatically accesses the game's main and UI camera frustums to check for visible characters or locations. If detected any of these game objects, the mod automatically logs their names for subsequent analysis. Additionally, our mod integrates with \textit{Steam's achievement system}\footnote{https://store.steampowered.com}, known for its comprehensive tracking capabilities. When an in-game achievement is unlocked, the mod automatically captures a screenshot, logging the current characters, locations, and the specific achievement. This ensures key game moments are documented for further analysis.

\par It is important to note that our mod is built with rigorous privacy safeguards. It neither interferes with the game's proprietary data nor collects, stores, or transmits any personal player information. It only captures in-game clues relevant to the study, ensuring both game integrity and privacy protection.

\subsubsection{Categorizing Six Elements and Providing Descriptions}
\par To enhance classification, we integrated GPT-4o with image interpretation capabilities. After capturing visual data, the LLM categorizes six specific game elements, generates descriptions, and extracts keywords, helping creators understand the clues (\Understanding{DR2}). This improves the accuracy and nuance of the classification, aligning with the predefined taxonomy. \xiyuan{To facilitate these tasks, we carefully and iteratively crafted prompts to ensure relevance and consistency. Details of the prompt engineering are provided in appendix~\ref{app:prompts}.}

\subsection{Interface}
\par Based on our design requirements and taxonomy, we developed an \xiyuan{open-source} interactive interface\footnote{\url{https://cluecart.github.io/ClueCart/}} to assist creators in analyzing and reconstructing game narratives using collected game clues. \xiyuan{Additionally, the interface includes a built-in system guide\footnote{\url{https://tasty-trick-667.notion.site/cluecart-guide-book}} to help users quickly get started.} In response to \textbf{RQ2}, we examine how these interface features support creators throughout their workflows. The interface is divided into two primary components, each tailored to help creators categorize and retrieve game clues and interpret game stories.

\subsubsection{Clue Categorization and Retrieval}
\par The left side of the interface (\cref{fig:cluecart}(A)) is designed to assist creators in categorizing, understanding, and retrieving game clues based on the hierarchical taxonomy. The interface is divided into three main sections: \textbf{\textit{Retrieval}} (\cref{fig:cluecart}(A1)) and \textbf{\textit{Classification}} (\cref{fig:cluecart}(A2)) on the left and \textbf{\textit{Recommended Elements}} (\cref{fig:cluecart}(A4)) on the right. Each section is specifically tailored to support different aspects of the creators' workflow, aligning with our design requirements (\Classification{DR1}-\Interpretation{DR4}).

\par The \textbf{\textit{Classification}} column on the left presents clues that have been processed and categorized according to the first level of the taxonomy. These are organized into three classifications: \textit{Character}, \textit{Location}, and \textit{Achievement}. Each acts as a tag applied to relevant clues (\cref{fig:cluecart}(A3)), providing a structured overview. This tagging system enables creators to quickly identify and group related clues, streamlining organization and retrieval (\Interpretation{DR4}).

\begin{figure*}[h]
  \centering
  \includegraphics[width=0.9\linewidth]
  {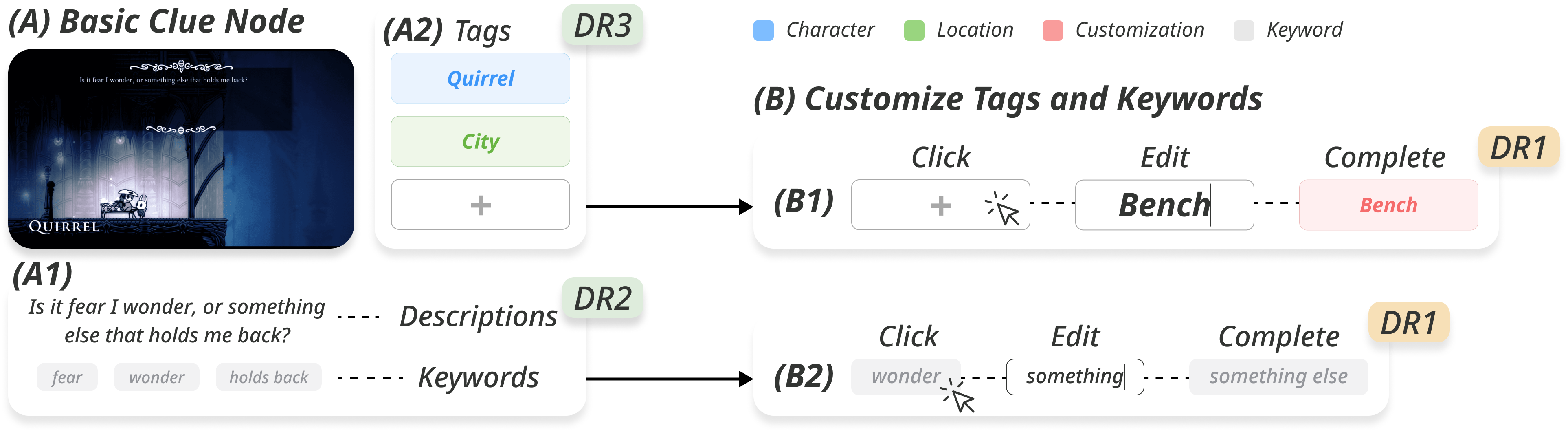}
   \caption[]{\xiyuan{Clue Node Design and Features: (A) Node design with (A1) tag list and (A2) descriptions and extracted keywords. (B) Tag / keyword customization via (B1-2) clicking and editing.}}
  \label{fig:node_revise}
\end{figure*}

\begin{figure*}[h]
  \centering
  \includegraphics[width=\linewidth]
  {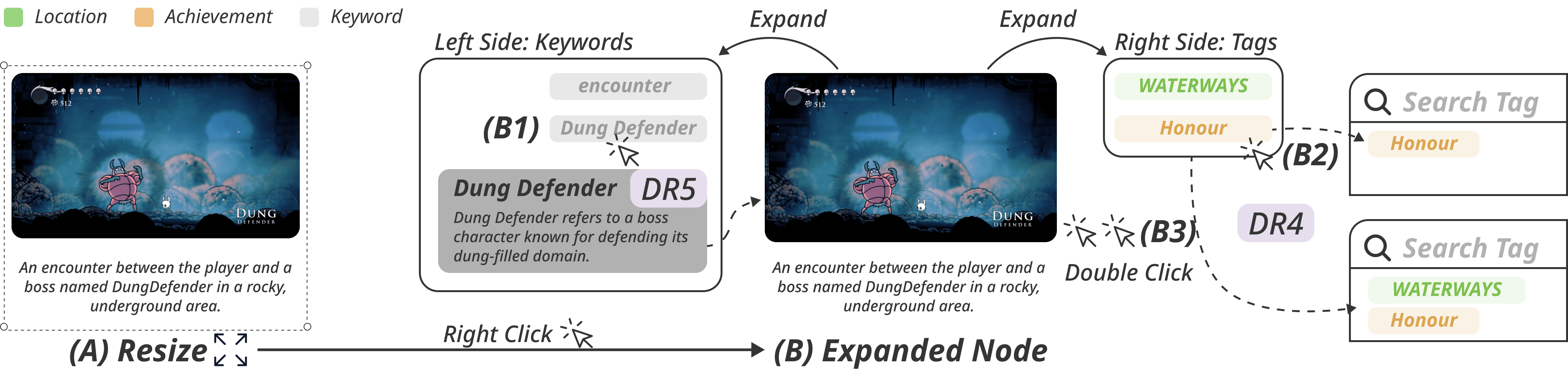}
  \caption[]{Interactions of a clue node. (A) Drag the bounding box to resize the node. (B) Expand the node by right clicking. (B1) Click on any keyword on the left to search for its real-world meaning. (B2) Click on a tag on the right to include it in the search query. (B3) Double-click on the clue node to add all associated tags to the search.}
  \label{fig:node}
\end{figure*}

\par The right column, labeled \textbf{\textit{Recommended Elements}}, further organizes game clues into six elements: \textit{Cut-scene}, \textit{Related Character}, \textit{Environment}, \textit{Artifact}, \textit{Dialogue}, and \textit{Text}. Each group includes relevant clues (\cref{fig:node_revise}), accompanied by descriptions, keywords (\cref{fig:node_revise}(A1)), and tags (\cref{fig:node_revise}(A2)). Descriptions and keywords are automatically generated for each clue to help creators quickly grasp its content and context, aligning with \Understanding{DR2}. Tags next to each clue highlight clue relationships, directly aiding in understanding these connections as outlined in \Understanding{DR3}.

\par To enhance usability, we integrated several features that support customized classification and efficient retrieval. The \textbf{\textit{Retrieval}} function in the upper-left corner (\cref{fig:cluecart}(A1)) offers robust search capabilities via tags. Creators can drag tags to adjust the priority of recommended node tags, prompting more personalized suggestions (\Interpretation{DR4}). In the \textbf{\textit{Recommended Elements}} section (\cref{fig:cluecart}(A4)), \xiyuan{each clue node incorporates two customization features for tags and keywords (\Classification{DR1}). A ``+ tag'' option (\cref{fig:node_revise}(B1)) lets users create and apply custom tags for custom classification. Additionally, as the LLM performance in word extraction can be inconsistent, creators can edit keywords to better summarize the descriptions (\cref{fig:node_revise}(B2)).} Furthermore, a sorting feature that arranges clues by temporal order and tag relevance (\cref{fig:cluecart}(A5)) helps creators view clues within context and enhancing their understanding of clue relationships (\Understanding{DR3}).

\subsubsection{Story Interpretation}
\par The \textbf{\textit{Interpretation}} panel (\cref{fig:cluecart}(B)) provides creators with a flexible workspace for organizing and analyzing collected clues, facilitating the reconstruction of the story's narrative. Creators can drag clues directly from the \textit{\textbf{Recommended Elements}} column onto the canvas, where they can engage with and manipulate these clues using various features designed to support reasoning and interpretation.

\par \textit{\textbf{Interactive Node Features.}} When creators interact with individual clue nodes, several features enhance their analysis. Clicking on a node with the left mouse button displays a bounding box (\cref{fig:node}(A)), allowing creators to resize the clue as needed. Right-clicking on a node reveals keywords on the left and corresponding tags on the right (\cref{fig:node}(B)), providing quick access to contextual information. Clicking on a keyword (\cref{fig:node}(B1)) prompts \xiyuan{(see appendix~\ref{app:prompts4rec})} \textit{ClueCart} to use an LLM to search for its real-world meaning, placing the result in a new node on the canvas, thus seamlessly integrating external information into their analysis (\Interpretation{DR5}). Additionally, clicking on a tag (\cref{fig:node}(B2)) initiates a search in the \textit{\textbf{Retrival}} section, and double-clicking a clue node (\cref{fig:node}(B3)) adds all its tags to this field, triggering the system to recommend related clues from the \textbf{\textit{Recommended Elements}} list, in line with \Interpretation{DR4}.

\par \textbf{\textit{Establishing Relationships Between Nodes.}} Beyond individual node interactions, the interface facilitates reasoning about the relationships between clues through a node-linking feature (\cref{fig:cluecart}(B1)). This functionality enables creators to draw connections between different nodes, visually mapping the relationships they identify. By double-clicking on a link, creators can add textual annotations, which enhances their analysis of how clues interrelate and contributes to a more thorough narrative reconstruction.

\par \textbf{\textit{Customization and Organizational Features.}} To further enhance the reasoning process, the \textit{Interpretation} panel offers customization options such as group nodes and text nodes. Group nodes (\cref{fig:cluecart}(B2)) enable creators to cluster related clues, ensuring a clear and organized structure that simplifies the navigation of complex narratives. Text nodes (\cref{fig:cluecart}(B3)) allow creators to document their thoughts, hypotheses, and insights directly on the canvas, facilitating ongoing reflection and interpretation as they refine their understanding of the story.

\section{Evaluation}
\label{sec: evaluation}
\par We conducted a between-subjects study (N=$40$) to evaluate \textit{ClueCart} against the baseline tool, \textit{Miro}.

\subsection{Baseline and Data Settings}
\par To evaluate the impact of \textit{ClueCart} on creators' story interpretation efficiency (\textbf{RQ3}), we used \textit{Miro} as the baseline for comparison. \xiyuan{\textit{Miro} was selected due to its widespread use among story interpreters, thanks to its flexibility, ease of use, and robust support for visual storytelling~\cite{berendsen2018digital,groshans2019digital,radics2021methodological,klein2023reimagining}. These features made \textit{Miro} an ideal baseline for evaluating interactive story interpretation across different media.}

\par In our study, we selected two games utilizing indexical storytelling: \textbf{\textit{Hollow Knight}}\xiyuan{~\cite{gameHollowKnight}} and \textbf{\textit{Do Not Feed the Monkeys (DNFTM)}}\xiyuan{~\cite{gameDNFTM}}.
% \footnote{\url{https://store.steampowered.com/app/658850/Do_Not_Feed_the_Monkeys/}}
The choice of these games was informed by feedback from our Formative Study (\cref{sec:formative study}), which indicated that complex gameplay in previous games hindered narrative interpretation. To address this, we included \textit{Hollow Knight} for its rich environmental storytelling and added \textit{DNFTM}, a puzzle-based game with simpler controls, to minimize cognitive load and enhance participants' focus on story analysis. To maintain consistency and control across participants, we pre-collected all game-related clues instead of asking participants to collect independently. Given the extensive time required to uncover key narrative elements in \textit{Hollow Knight}, gathering clues in advance ensured that all participants had access to the same data, facilitating a fair comparison. The collected clues, adhering to our standard gameplay procedure, included all six elements in the taxonomy, ensuring both input consistency and ecological validity.

\subsection{Participants}
\par We recruited $40$ participants ($18$ females, $22$ males, aged $20$-$29$) through social media channels (~appendix~\ref{app:user_study_participants}). All participants had prior experience with discussing and interpreting game stories: $19$ through video platforms, $1$ on online forums, and $20$ through a combination of both. In addition, one participant is a contributor of a game wiki. \xiyuan{These players, who are also creators of game story interpretations,} were randomly assigned to one of two groups of $20$ people each: one group analyzed \textit{Hollow Knight} and the other \textit{DNFTM}. Each group was further split into two subgroups of $10$, with one subgroup using \textit{Miro} for story interpretation and the other using \textit{ClueCart}. This between-subjects design was employed to address the substantial disparity in gameplay difficulty between \textit{Hollow Knight} and \textit{DNFTM}. To assure familiarity with the game stories and reduce potential learning curve, all participants had previously played through the games before the study. Participants were compensated $\$10$ per hour.

\subsection{Evaluation Dimensions}
\label{sec: eval_dimension}
\par \xiyuan{We adopted three key dimensions for quantitative evaluation: \textbf{\textit{Perception}}, \textbf{\textit{Features}}, and \textbf{\textit{Story Interpretation Assessment}}, as shown in ~appendix~\ref{app:dimension}.}

\par \xiyuan{The \textbf{\textit{Perception}} dimension captured participants' subjective experiences, focusing on usability and creativity support. Two widely adopted metrics were used: \textit{System Usability Scale} (SUS)~\cite{bangor2008empirical} and \textit{Creativity Support Index} (CSI)~\cite{cherry2014quantifying}. The SUS assessed the overall usability of the tools, while the CSI focused on how well the tools supported participants' creative processes. Both of them are commonly used in evaluation of creativity support tools~\cite{frich2024measures,tchemeube2023evaluating}, offering reliable insights into participants' subjective impressions and satisfaction with the tool. }

\par \xiyuan{The \textbf{\textit{Features}} dimension evaluated \textit{ClueCart}'s specific functionalities, aligning with the categories, challenges (Cs) and design requirements (DRs) outlined in \cref{tab:challenges}. This dimension was further divided into three aspects: \textit{taxonomy}, corresponding to Classification \& Organization (C1, C2; \Classification{DR1}); \textit{functionality}, addressing Clue Understanding (C3, C4; \Understanding{DR2}, \Understanding{DR3}) and Story Interpretation (C5, C6; \Interpretation{DR4}, \Interpretation{DR5}) and \textit{interaction}, reflecting participants' impressions of \textit{ClueCart}'s overall interactions. This dimension is specifically evaluated by \textit{ClueCart} group. }

\par \xiyuan{The \textbf{\textit{Story Interpretation Assessment}} dimension focused on participants' performance in reconstructing narratives, with two key metrics: \textit{task completion time}, measuring the time taken to complete the task, and \textit{peer scoring}, which used a concept map evaluation checklist\footnote{\url{https://www.morgan.edu/Documents/ADMINISTRATION/DIVISIONS/IT/ConceptMap-EvalChecklist.pdf}} to assess outputs based on clarity, organization, and logical structure. Although \textit{Miro} outputs may not conform to a traditional concept map format, this framework was effective for assessing key aspects such as clarity, organization, and logical structure. This standardized evaluation ensured that consistent assessment of how accurately participants represented the game's narratives.}

\par \xiyuan{To complement these quantitative methods, we conducted \textit{\textbf{semi-structured interviews}} to gather qualitative feedback on participants' experiences, providing deeper insights into strengths, weaknesses, and improvement areas. By integrating quantitative and qualitative methods, this evaluation ensured a rigorous and holistic assessment of \textit{ClueCart}'s usability, functionality, and narrative support capabilities.}

\subsection{Study Setup}
\par To assess the feasibility and effectiveness of our study procedure, two members of our research team conducted a pilot study, each focusing on one of the games. This process helped refine the study design and ensure that tasks were appropriately structured to align with the narrative style of each game.

\par Prior to each session, participants were provided with a consent form, background information, and an overview of the research goals to ensure they were well-informed. Additionally, we distributed a pre-collected game material package, which included detailed explanations of the clue collection process, data format, and organizational structure. This preparation facilitated a smooth progression of the study. For participants in the \textit{Miro} group, we provided a tutorial on how to use the tool, while those in the \textit{ClueCart} group received a system guide and instructional video to help them understand the interface.

\subsection{Study Procedure}
\par The formal study was divided into four parts: \textit{\textbf{introduction}}, \textit{\textbf{story interpretation}}, \textit{\textbf{evaluation and peer scoring}}, and \textit{\textbf{feedback}}.

\subsubsection{Introduction ($20$ min for \textit{Miro} group, $25$ min for \textit{ClueCart} group)}
\par In the overall introduction part ($15$ min), we explained the whole study procedure, offering a sequential overview and estimated duration for each segment. We introduced the research background and objectives, illustrating how to analyze the game narratives from three key perspectives: \textit{plot analysis}, \textit{character relationships}, and \textit{character portrait}. We also reviewed the game clues provided the prior day, outlining their collection during gameplay and elucidating the selection criteria to guarantee the inclusion of all relevant narrative elements, so providing participants with a thorough comprehension of the dataset.

\par Following this, tool-specific orientations were given to each subgroup to ensure participants got only material relevant to their designated tool. The \textit{Miro} group received a brief introduction ($5$ min) on the tool's visual collaboration features, emphasizing its application in narrative analysis mapping. The \textit{ClueCart} group was given a comprehensive review ($10$ min) of the two-level taxonomy derived from the formative study, along with a demonstration of how to use the \textit{ClueCart} interface to organize and connect narrative fragments. This preparation ensured all participants were familiar with their respective tools before starting the task.

\subsubsection{Story Interpretation ($60$ min)}
\par Following the introduction, participants proceeded to the story interpretation task, where they analyzed game clues from the perspectives of \textit{plot analysis}, \textit{character relationships}, and \textit{character portrait}. The time taken by each participant to complete the task was recorded and subsequently used to assess the efficiency of the tools in supporting narrative interpretation.

\subsubsection{Evaluation and Peer Scoring ($15$ min)}
\par \xiyuan{After completing the task, participants filled out a $7$-point Likert scale questionnaire, assessing the \textit{Perception} dimension, including SUS and CSI, based on their tool experience.
In particular, we incorporated questions about specific \textit{Features} of \textit{ClueCart}, focusing on taxonomy, functionality, and interactions for the \textit{ClueCart} group.
For participants' \textit{Story Interpretations}, we recorded their completion time and included peer scoring in the questionnaire to evaluate clarity, organization, and logical structure of the interpretations.}

%The questionnaire employed the \textit{System Usability Scale} (SUS)~\cite{bangor2008empirical}, \textit{Creativity Support Index} (CSI)~\cite{cherry2014quantifying}. The SUS assessed the overall usability of the tools, while the CSI focused on how well the tools supported participants' creative processes. 
% To evaluate the participants' story interpretations, we employed the overall section of a concept map \textit{evaluation checklist}\footnote{\url{https://www.morgan.edu/Documents/ADMINISTRATION/DIVISIONS/IT/ConceptMap-EvalChecklist.pdf}}. Although \textit{Miro} outputs may not conform to a traditional concept map format, this framework was effective for assessing key aspects such as clarity, organization, and logical structure. Both tools required participants to interpret complex stories, organize clues, and present them coherently. This standardized evaluation ensured that consistent assessment of how accurately participants represented the game's narratives. \xiyuan{All questions were rated on a $7$-point Likert scale.}

\subsubsection{Feedback ($20$ min)}
\par Following the scoring phase, we conducted semi-structured interviews ($20$ min) with each participant to gather detailed feedback on their experience with the tool. During the interviews, participants were first asked to describe their ideation process throughout the task, followed by their overall experience with the tool they used. We then revisited the taxonomy introduced during the study to gather feedback and suggestions for enhancement. Finally, we examined participants' perspectives on the future of game story interpretation. These qualitative insights complemented the quantitative data from the questionnaire, offering a deeper understanding of the strengths and limitations of both \textit{Miro} and \textit{ClueCart} in facilitating interactive story interpretation.

\section{Results}
% \par We collected participants' ratings on usability, creativity support, peer narrative interpretation outcomes, and task completion duration. Following this, semi-structured interviews were conducted to gather qualitative insights into their ideation processes, tool experiences, and suggestions on improvements.

\par \xiyuan{We collected participants' ratings on the three main dimensions mentioned in \cref{sec: eval_dimension}, including \textbf{\textit{Perception}}, \textbf{\textit{Features}} and \textbf{\textit{Story Interpretation Assessment}}. All ratings were collected using a $7$-point Likert scale. }

\par \xiyuan{Quantitative data for perception and story interpretation assessment was analyzed using the Mann-Whitney U Test~\cite{mcknight2010mann} to compare \textit{ClueCart} with the baseline tool, \textit{Miro}. Ratings for features of \textit{ClueCart} were summarized descriptively without comparative statistical analysis.} For the qualitative analysis, interview transcripts were coded using inductive thematic analysis~\cite{braun2012thematic}, with two researchers cross-checking to ensure consistency. This mixed-methods approach allowed us to identify key themes and patterns in participants' feedback, complementing the quantitative findings.

% \par The quantitative data, assessed via a $7$-point Likert scale for usability and creativity support, was analyzed through the Mann-Whitney U Test~\cite{mcknight2010mann} to compare the tools. For the qualitative analysis, interview transcripts were coded using inductive thematic analysis~\cite{braun2012thematic}, with two researchers cross-checking to ensure consistency. This combined approach allowed us to identify key themes and patterns in participants' feedback, complementing the quantitative findings.

\subsection{Perception of \textit{ClueCart}}
\par \xiyuan{The perception of \textit{ClueCart} was assessed across multiple aspects, including system usability (SUS) and creativity support (CSI), with results compared with baseline \textit{Miro}. Statistical analyses determined the significance of differences between the two tools, with a threshold set at $p<0.05$. Specific levels of significance are indicated in \cref{tab:sus_csi}, where * represents $p<0.05$, ** represents $p<0.01$, and *** represents $p<0.001$. Means ($M$) and standard deviations ($SD$) are reported for each metric. The results demonstrated statistically significant improvements over the baseline system across these dimensions. Detailed questions for these aspects are provided in appendix~\ref{app:perception_questions}. }

\begin{table*}[h]
    \caption{The statistical user feedback with \textit{Baseline} and \textit{ClueCart}, where the $p$-values (*:$p<.050$, **:$p<.010$, ***:$p<.001$) is reported.}
    \centering
    \begin{tabular}{llcccc}
        \toprule
        \multirow{2}{*}{\textbf{Categories}} & \multirow{2}{*}{\textbf{Factors}} & \small{\textbf{Baseline}} &\small{\textbf{ClueCart}} & \multirow{2}{*}{\textbf{$P$-value}} & \multirow{2}{*}{\textbf{Sig.}} \\
        &  & \small{\textbf{Mean/S.D.}} &\small{\textbf{Mean/S.D.}} & &\\
        \midrule
        \multirow{6}{4cm}{System Usability Scale} 
        & Easy to use & $5.15/0.70$ & $6.10/0.59$ & $1.79 \times 10^{-4}$ & ***\\
        & Functions & $4.07/1.07$ & $6.15/0.64$ & $6.92 \times 10^{-7}$ & ***\\
        & Quick to learn & $4.87/0.92$ & $6.12/0.83$ & $2.80 \times 10^{-4}$ & ***\\
        & Learning curve & $3.20/1.00$ & $1.70/0.80$ & $8.00 \times 10^{-5}$ & ***\\
        & Frequency & $3.65/1.24$ & $6.07/0.65$ & $7.95 \times 10^{-7}$ & ***\\
        & Confidence & $4.57/1.20$ & $6.18/0.64$ & $2.45 \times 10^{-5}$ & ***\\
        \midrule
        \multirow{5}{4cm}{Creativity Support Index} 
        & Enjoyment & $4.05/1.32$ & $6.10/0.86$ & $1.50 \times 10^{-5}$ & ***\\
        & Exploration & $3.93/1.29$ & $6.28/0.63$ & $4.26 \times 10^{-6}$ & ***\\
        & Results Worth Effort & $4.12/1.63$ & $6.02/1.04$ & $4.20 \times 10^{-4}$ & ***\\
        & Expressiveness & $3.67/1.18$ & $5.92/1.02$ & $1.50 \times 10^{-5}$ & ***\\
        & Immersion & $3.55/1.19$ & $6.12/0.86$ & $6.92 \times 10^{-7}$ & ***\\
        \bottomrule
    \end{tabular}
    \label{tab:sus_csi}
\end{table*}

\subsubsection{System Usability}
\par The system usability~\cite{bangor2008empirical} of \textit{ClueCart} was assessed through various measures including \textit{ease of use}, \textit{functionality}, \textit{learning curve}, \textit{frequency of use}, and \textit{user confidence}, as summarized in \cref{tab:sus_csi} and appendix~\ref{app:perception}.

\par \xiyuan{Participants rated \textit{ClueCart} as significantly easier to use ($M = 6.10$, $SD = 0.59$) than the baseline system ($M = 5.15$, $SD = 0.70$). Similarly, functionality was evaluated as having more comprehensive and better-integrated features ($M = 6.15$, $SD = 0.64$) than the baseline system ($M = 4.07$, $SD = 1.07$). Learnability was also rated significantly higher for \textit{ClueCart} ($M = 6.12$, $SD = 0.83$) compared to the baseline system ($M = 4.87$, $SD = 0.92$). Additionally, \textit{ClueCart} was perceived to have a significantly lower learning curve ($M = 1.70$, $SD = 0.80$) than the baseline ($M = 3.20$, $SD = 1.00$), indicating its ease of adoption and user-friendliness. Participants also indicated that they would use \textit{ClueCart} more frequently ($M = 6.07$, $SD = 0.65$) than the baseline system ($M = 3.65$, $SD = 1.24$) and felt greater confidence while using it ($M = 6.18$, $SD = 0.64$) compared to the baseline system ($M = 4.57$, $SD = 1.20$). Detailed p-values for these comparisons are presented in \cref{tab:sus_csi}.}

% \par Participants rated \textit{ClueCart} as significantly easier to use ($M = 6.10$, $SD = 0.59$) compared to the baseline system ($M = 5.15$, $SD = 0.70$), with a $p$-value of $< 0.001$. In terms of functionality, \textit{ClueCart} was evaluated as having more comprehensive and better-integrated features ($M = 6.15$, $SD = 0.64$) compared to the baseline system ($M = 4.07$, $SD = 1.07$), with a $p$-value of $< 0.001$. Learnability was also rated significantly higher for \textit{ClueCart} ($M = 6.12$, $SD = 0.83$) compared to the baseline system ($M = 4.87$, $SD = 0.92$), with a $p$-value of $< 0.001$. Additionally, \textit{ClueCart} was perceived to have a significantly lower learning curve ($M = 1.70$, $SD = 0.80$) than the baseline ($M = 3.20$, $SD = 1.00$), indicating its ease of adoption and user-friendliness ($p$-value of $< 0.001$). Participants also indicated that they would use \textit{ClueCart} more frequently ($M = 6.07$, $SD = 0.65$) than the baseline system ($M = 3.65$, $SD = 1.24$), with a $p$-value of $< 0.001$. Furthermore, they reported greater confidence using \textit{ClueCart} ($M = 6.18$, $SD = 0.64$) compared to the baseline system ($M = 4.57$, $SD = 1.20$), with a $p$-value of $< 0.001$.

\subsubsection{Creativity Support}
\par As shown in \cref{tab:sus_csi} and appendix~\ref{app:perception}, the \textit{Creativity Support Index} (CSI)~\cite{cherry2014quantifying} for \textit{ClueCart} was evaluated across five factors: \textit{enjoyment}, \textit{exploration}, \textit{perceived effort}, \textit{expressiveness}, and \textit{immersion}. We excluded the inapplicable \textit{collaboration}
dimension.

\par \xiyuan{Participants rated \textit{ClueCart} significantly higher in enjoyment ($M = 6.10$, $SD = 0.86$) compared to the baseline system ($M = 4.05$, $SD = 1.32$). The system also encouraged greater exploration of creative avenues ($M = 6.28$, $SD = 0.63$) compared to the baseline system ($M = 3.93$, $SD = 1.29$). The perceived value of the results achieved with \textit{ClueCart} was also higher ($M = 6.02$, $SD = 1.04$) than the baseline ($M = 4.12$, $SD = 1.63$). In terms of expressiveness, \textit{ClueCart} received a higher rating ($M = 5.92$, $SD = 1.02$) than the baseline system ($M = 3.67$, $SD = 1.18$). Additionally, participants reported a significantly greater sense of immersion in the creative process ($M = 6.12$, $SD = 0.86$) with \textit{ClueCart} compared to the baseline system ($M = 3.55$, $SD = 1.19$). Overall, \textit{ClueCart} demonstrated significant improvements in creativity support across all the evaluated dimensions. Detailed p-values are presented in \cref{tab:sus_csi}.}

% \par Participants rated \textit{ClueCart} significantly higher in enjoyment ($M = 6.10$, $SD = 0.86$) compared to the baseline system ($M = 4.05$, $SD = 1.32$). The system also encouraged greater exploration of creative avenues ($M = 6.28$, $SD = 0.63$) compared to the baseline system ($M = 3.93$, $SD = 1.29$). The perceived value of the results achieved with \textit{ClueCart} was also higher ($M = 5.92$, $SD = 1.04$) compared to the baseline ($M = 4.12$, $SD = 1.63$). In terms of expressiveness, \textit{ClueCart} received a higher rating ($M = 5.91$, $SD = 1.02$) compared to the baseline system ($M = 3.67$, $SD = 1.18$). Additionally, participants reported a significantly greater sense of immersion in the creative process ($M = 6.12$, $SD = 0.86$) with \textit{ClueCart} compared to the baseline system ($M = 3.55$, $SD = 1.19$). Overall, \textit{ClueCart} demonstrated significant improvements in creativity support across all the evaluated dimensions.

\subsection{Features of \textit{ClueCart}}

\par \xiyuan{The features of \textit{ClueCart} were evaluated based on three key aspects: \textit{taxonomy}, \textit{functionality}, and \textit{interaction}. Each aspect focused on a specific feature of how participants engaged with the tool's capabilities during story interpretation tasks. Unlike perception, features were assessed descriptively rather than comparatively with baseline \textit{Miro}. The results are shown in \cref{tab:feature_ratings} while detailed questions corresponding to each feature aspect are provided in appendix~\ref{app:feature_questions}.}

\begin{table*}[ht]
    \caption{User ratings on \textit{ClueCart} features.}
    \centering
    \begin{tabular}{lllccc}
        \toprule
        \textbf{Categories} & \textbf{DR} & \textbf{Features} & \textbf{Mean} &\textbf{S.D.} \\
        \midrule
        \multirow{1}{*}{Taxonomy} 
        & \Classification{DR1} & Clue Categorization & 6.35 & 0.73  \\
        \midrule
        \multirow{4}{*}{Functionality} 
        & \Understanding{DR2} & Content Understanding & 6.30 & 0.78 \\
        & \Understanding{DR3} & Relationship Understanding & 6.10 & 0.77 \\
        & \Interpretation{DR4} & Retrieving Clues & 6.30 & 0.71 \\
        & \Interpretation{DR5} & New Insights from External Information & 6.35 & 0.79 \\
        \midrule
        \multirow{1}{*}{Interaction} 
        &  & Overall System Interaction & 6.20 & 0.75 \\
        \bottomrule
    \end{tabular}
    \label{tab:feature_ratings}
\end{table*}

\subsubsection{Taxonomy: Classification $\&$ Organization}
\par The taxonomy feature in \textit{ClueCart}, particularly its clue categorization function, received high praise from participants, with an average score of $6.35$ ($SD = 0.73$). This aligns with \Classification{DR1}, suggesting that the system effectively mirrors users' natural organization of story clues. Most participants ($37$ out of $40$) expressed strong approval of the taxonomy. P$1$ mentioned, ``\textit{This categorization aligns with how I typically organize story clues in games.}'' \xiyuan{However, not all feedback was positive. Further analysis revealed that while the taxonomy was appreciated by most, some participants found it challenging to use. P$23$ noted that ``\textit{understanding and adapting to this taxonomy takes some time—there's a bit of a learning curve}'' and suggested integrating a detailed tutorial to ease adoption. P$32$ mentioned that while the taxonomy was helpful for structured analysis, it sometimes hindered creative exploration, stating, ``\textit{I feel constrained by the classification when trying to think outside the box.}'' P$3$ and P$10$ recommended that future iterations explore quest-based clue categorization to align with built-in quest structures in some games, further enhancing the experience. These mixed responses highlight that while the taxonomy provides a useful framework, it can be difficult to adapt when game narratives or clues do not neatly fit predefined categories. This underscores the need for a more flexible, user-customizable taxonomy, particularly at the classification level, which remains a key limitation of the current design.}

% P$10$ suggested that since some games feature built-in task lists, future iterations could explore clue categorization based on these task structures. Similarly, P$3$ recommended incorporating task-based clue categorization in future updates to further enhance the experience. However, not all feedback was positive. P$23$ noted that while helpful, ``\textit{understanding and adapting to this taxonomy takes some time—there's a bit of a learning curve}'', \xiyuan{She suggested integrating a more detailed tutorial to ease adoption. Further analysis of participant feedback revealed that while most appreciated the taxonomy, some found the it challenging, especially when mapping complex or ambiguous clues. For instance, P32 expressed that while the taxonomy was useful for structured analysis, it sometimes hindered creative exploration, noting, ``\textit{I feel constrained by the classification when trying to think outside the box.}'' These mixed responses underscore that while the taxonomy offered a helpful framework, some users faced challenges when game narratives or clues did not neatly align with predefined categories. This highlights the need for a more flexible, user-customizable taxonomy, particularly at the Classification level, which remains a key limitation in the current design.}

\subsubsection{Functionality: Clue Understanding}
\par The average score for content understanding was $6.30$ ($SD = 0.78$), showing that participants found the system highly effective in helping them grasp story clues, aligning with \Understanding{DR2}. Participants noted that summarizing visual clues or extracting keywords significantly improved their understanding of clues. P$15$ noted, ``\textit{Interpreting images can be exhausting, especially with large amounts of clues.}'' P$38$, who worked on the \textit{DNFTM} project, mentioned, ``\textit{In a text-heavy game like DNFTM, manually reviewing each clue is laborious. Summaries and keyword extraction make it easier to quickly grasp each clue's content.}'' Similarly, P$18$ added, ``\textit{When working on the presidential election story in DNFTM, the information was often scattered across newspapers and dialogues. Without the system summarizing the content, it would have been overwhelming.}''

\par The average score for relationship understanding was $6.10$ ($SD = 0.77$), indicating that \textit{ClueCart} effectively aids users in identifying connections between clues, consistent with \Understanding{DR3}. P$15$ noted, ``\textit{The tag system makes it easier to track clue connections. Visualizing relationships helped me see links I hadn't noticed before}'' However, some participants offered constructive feedback, with six (P$13$, P$15$, P$14$, P$19$, P$32$, P$39$) suggesting room for improvement. For instance, P$39$ said, ``\textit{The relationship view feels a bit manual... I first look for clues with similar tags, then check for new ones layer by layer. It works for about $80\%$ of cases, but more automation could help. If the system suggests relationships based on visuals, text, and tags after selecting two clues, that'll really boost my workflow and give me new insights.}'' \xiyuan{Their feedback points to opportunities for improving understanding efficiency.}

\subsubsection{Functionality: Story Interpretation.}
\par In terms of retrieving clues, \textit{ClueCart} scored $6.30$ ($SD = 0.71$), demonstrating its effectiveness in helping users quickly find necessary ones, aligning with \Interpretation{DR4}. Most participants in the \textit{ClueCart} group ($19$ out of $20$) found the filtering and retrieval features highly beneficial. P$17$ noted, ``\textit{It's really convenient to pull up what I need.}'' Improvements still remained. P$16$ suggested improving the sorting and recommendation by incorporating techniques like cosine similarity. Furthermore, P$36$, along with P$12$, P$34$, and P$37$, recommended adding a keyword search function within the retrieval process to further boost efficiency. This suggestion reflects a strong user demand for more advanced retrieval options to streamline content discovery and make the process even more intuitive.

\par The new insights from external information dimension achieved an average score of $6.35$ ($SD = 0.80$), showing that participants found the external search feature valuable for gaining new insights, supporting \Interpretation{DR5}. Those who used the feature extensively provided positive feedback. P$17$ highlighted how the feature enriched her experience with \textit{Hollow Knight}, explaining, ``\textit{Sly's name—`cunning' or `clever'—matches his character in the game. That deepened my understanding of him.}'' Similarly, P$35$ found the feature immersive while working on \textit{DNFTM}, remarking, ``\textit{Walker's slogan, `make the country glorious again', was clearly a nod to Trump. The system picked up on this and linked it to real-world context, making the narrative more immersive for me.}'' However, some participants ($6$ out of $20$) explored this feature only briefly, suggesting that it might not have been central to their workflow. P$39$ remarked, ``\textit{Most keywords were descriptive… I didn't really get the key points.}'' He suggested that the feature could be further enhanced by evolving from manual exploration to an automated system that identifies and highlights key insights for the user.

\subsubsection{Overall System Interaction}
\par The overall system interaction received an average score of $6.20$ ($SD = 0.75$), indicating that participants found the system effective for organizing clues and facilitating story interpretation tasks. Many participants praised the interpretation panel, particularly its node grouping and arrow functionality, which supported their reasoning process. The drag-and-drop feature, highlighted by P$12$, P$17$, P$32$, P$35$ for its engaging and interactive nature. P$17$ and P$31$ appreciated the intuitiveness and efficiency of the keyword search function, while the tagging system in the clue library was valued for it quick organization capabilities.

\par Despite these strengths, participants suggested areas for improvement. Arrows were difficult to select and delete individually, and P$31$ recommended a bulk selection feature. Text node interactions posed issues for P$33$ and P$12$, who found resizing and moving text unclear and limited to corner adjustments. P$32$ pointed out the need for better layering control for text and images, as text often overlapped visuals, affecting organization. P$19$ also noted that clue retrieval became challenging with large datasets, suggesting improvements for better search functionality. Overall, participants found the system intuitive and helpful, particularly the interpretation panel and search features, though improvements in arrow usage, text box interaction, layering, and clue retrieval were recommended to enhance workflow efficiency.

\subsection{Story Interpretation Assessment}
\par \xiyuan{The assessment of story interpretation using \textit{ClueCart}, compared to \textit{Miro}, focused on three main aspects: completion time, quality, and result comparison. They were analyzed to evaluate how \textit{ClueCart} influenced the efficiency and effectiveness of the story interpretation process. Detailed questions are provided in appendix~\ref{app:result_questions}.}

\begin{figure*}[h]
  \centering
  \includegraphics[width=\linewidth]
  {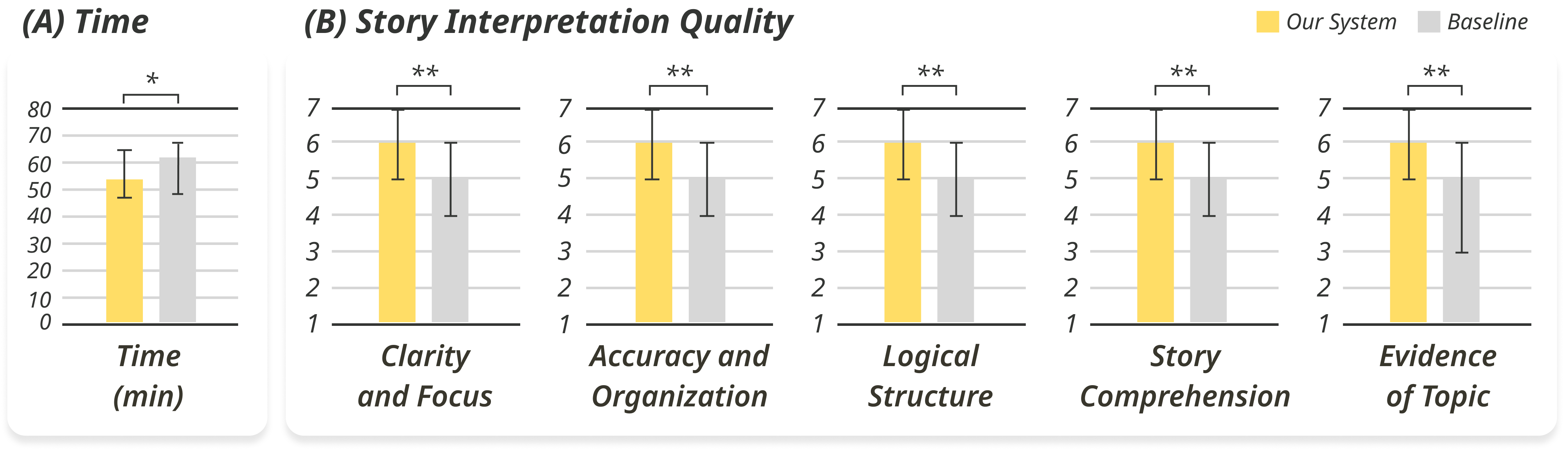}
  \caption[]{Story Interpretation Assessment. (A) Completion Time \xiyuan{of \textit{ClueCart} is shorter than baseline. (B) Story Interpretation Quality \xiyuan{using \textit{ClueCart} shows significant improvement compared to the baseline.}}}
  \label{fig:story_assess}
\end{figure*}

\begin{figure*}[h]
  \centering
  \includegraphics[width=\linewidth]
  {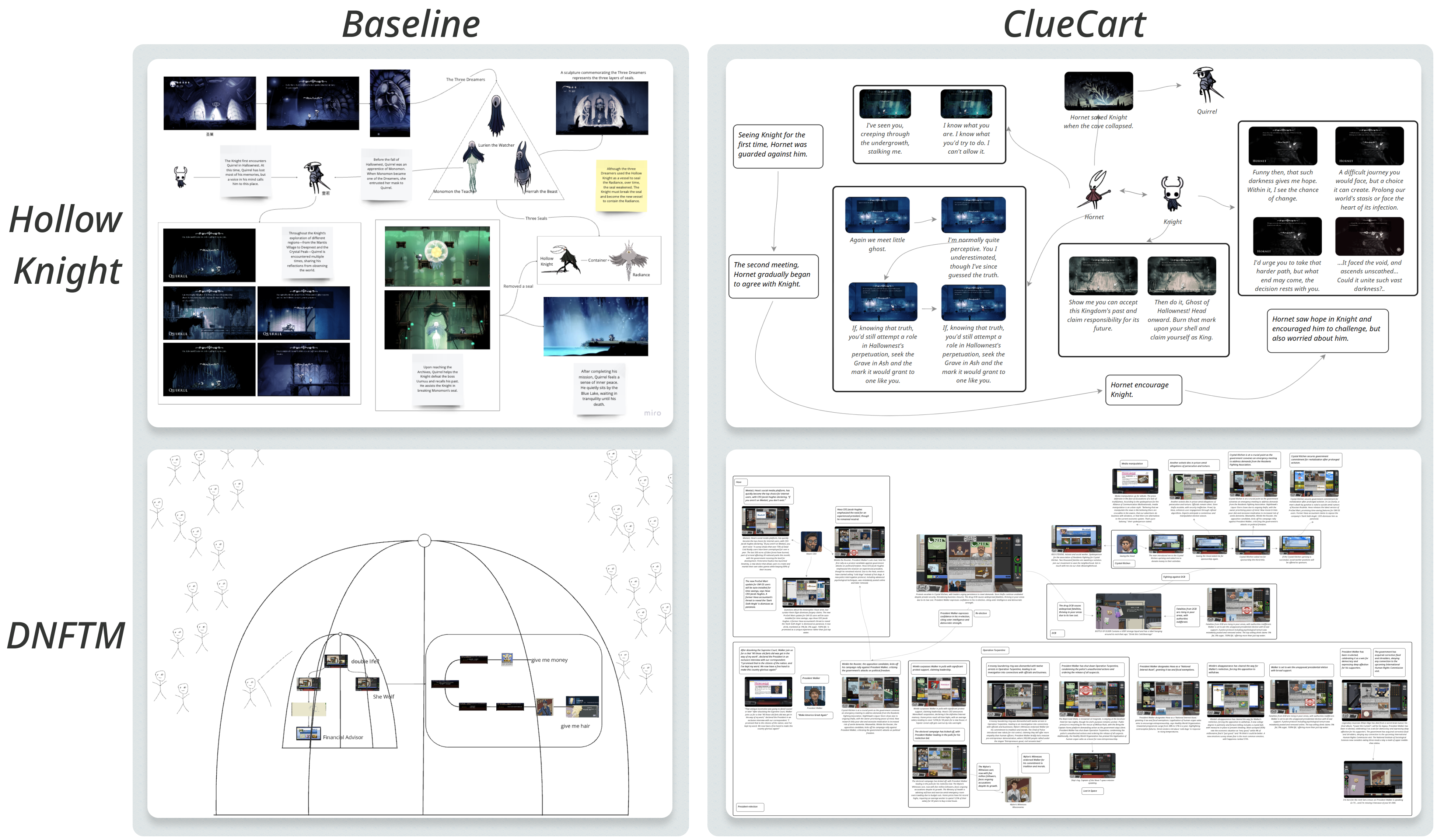}
  \caption[]{Story interpretation results of high average scores. \xiyuan{The presentations from the baseline group were more stylistically varied, while outputs of \textit{ClueCart} emphasized logical relationships between clues.}}
  \label{fig:story_results}
\end{figure*}

\subsubsection{Completion Time}
\par In terms of completion time (\cref{fig:story_assess} (A)), \textit{ClueCart} facilitated faster creative outputs, achieving a median time of $53$ minutes compared to $62.5$ minutes for the baseline system. Additionally, \textit{ClueCart} exhibited more consistent performance, with first and third quartile times of $47.25$ and $65.25$ minutes, respectively, compared to $49.5$ and $68.5$ minutes for the baseline. These results suggest that \textit{ClueCart} not only sped up task completion but also delivered more reliable performance across users.

\subsubsection{Quality}
\par In terms of quality (\cref{fig:story_assess} (B)), users rated their peers' work more favorably across all five dimensions---\textit{Clarity and Focus}, \textit{Accuracy and Organization}, \textit{Logical Structure}, \textit{Story Comprehension}, and \textit{Evidence of Topic }---when using \textit{ClueCart}. \textit{ClueCart} consistently achieved higher scores at all quartile levels, with a median score of $6$, compared to $5$ for the baseline. Notably, in the \textit{Evidence of Topic} category, \textit{ClueCart} showed significant improvement, with a first quartile score of $5$ compared to $3$ for the baseline. These results suggest that \textit{ClueCart} not only expedited task completion but also delivered superior-quality creative outputs compared to the baseline.

\subsubsection{Result Comparison}
\par Story interpretation results with high average scores are presented in \cref{fig:story_results}. In terms of flexibility, the Miro group's presentations were more stylistically varied. For example, in \textit{Hollow Knight}, P$5$ represented the relationships between the three \textit{Dream Keepers} using a triangle, while in \textit{DNFTM}, P$21$ used a ``cage'' metaphor to convey the layers surrounding the game world, highlighting its unique atmosphere. In contrast, \textit{ClueCart} outputs focused on logical relationships and completing clues, with users identifying more clues than the \textit{Miro} group. This suggests that \textit{ClueCart} encourages a more structured, systematic approach to story interpretation, prioritizing thoroughness and logical coherence over \xiyuan{creative expression}.

\section{Discussion}
\par In this section, we begin by comparing data collection methods, focusing on game mods and multi-modal classification models. We then explore \textit{ClueCart}'s impact on the \textit{Fan Creation Community}. Finally, we summarize design implications to guide future efforts in enhancing the efficiency, completeness, and creativity of story interpretation for creators.

\subsection{Exploration of Data Collection Techniques}

\par This study is grounded in the collection of raw materials from a variety of games. During the data collection process, we explored two methods: multi-modal classification models and in-game mods.

\par Multi-modal models can automate categorization using diverse data inputs, but they require large, labeled datasets for effective training. The variation in character appearances, locations, and achievements across games makes generalization difficult. Moreover, retraining these models for each game is costly and time-consuming, limiting their practicality in dynamic creative workflows. In contrast, in-game mods provide a more practical solution by enabling real-time extraction and categorization of game elements. This method integrates smoothly into workflows, capturing game-specific data during interaction for highly accurate classification. While mod development faces challenges like engine compatibility, our adaptable design principles mitigate these issues, allowing the mod to scale efficiently across different gaming environments.

\subsection{Fostering Fan Creation Community}

\subsubsection{Enhancing Content Appeal and Engagement.}
\par For video creators, maintaining content relevancy and fostering audience engagement are essential~\cite{zhou2016youtube,figueiredo2011tube}. Content development often involves complex tasks such as evaluating clues and constructing narratives, which can be daunting ($32$ out of $40$). \textit{ClueCart} offers features like clue retrieval, recommendations, and linking in-game elements to real-world meanings, enabling creators to discover unique insights (P$17$, P$35$). This helps craft unexpected yet rational narrative twists that captivate viewers, boosting engagement and content virality (P$32$). Additionally, the concept maps produced by \textit{ClueCart} can serve as supplementary resources for videos and forum discussion~\cite{10.1007/978-3-030-29384-0_20}, helping audiences better understand complex story connections (P$12$, P$18$, P$36$). These maps not only enhance content sophistication but also provide a structured way for audiences to follow along. Furthermore, \textit{ClueCart} empowers creators to present varied interpretations of a single story, encouraging vigorous discourse among viewers in the comments, enhancing interaction and promoting deeper engagement (P$39$, P$40$). This exchange of ideas enriches the content's depth and fosters the video's popularity and community discourse.

\subsubsection{Lowering Knowledge Barrier Through Story Interpretation}
\par As a new kind of art form~\cite{smuts2005video}, video games not only serve as a medium of entertainment but also possess significant cultural dissemination potential~\cite{chen2013video}. For instance, the \textit{God of War}~\cite{gameGOW} series introduces players to Norse mythology~\cite{tosi2022classical}, while \textit{Black Myth: Wukong}~\cite{gameWukong} is based on China's \textit{Journey to the West}~\cite{cheng2015journey}, showcasing rich cultural heritage~\cite{mao2024study}. Similarly, \textit{Assassin's Creed}~\cite{gameAssassin} incorporates diverse cultural and historical backgrounds across its series~\cite{seif2008assassin,politopoulos2019history,shaw2015tyranny}, making games a bridge for global cultural exchange. By embedding real-world culture and history into game narratives, creators can lower the knowledge barriers for players. Those familiar with specific cultural references often form deeper emotional connections with the game content, while others can acquire new cultural knowledge and historical insights through the narrative. This approach not only enhances the immersive experience for players but also positions video games as a potential platform for education and cross-cultural exchange~\cite{soyoof2018video,cwil2020cross}.

\subsubsection{Expanding Storytelling Beyond Games}
\par \textit{ClueCart} is not limited to game content but also has the potential to support other storytelling media, such as novels and films. This multi-media compatibility provides creators with a broader range of applications. For instance, creators can automatically categorize and manage key plot points, character relationships from novels, or visual symbols and narrative clues from films through \textit{ClueCart}. This integration allows different narrative clues to be organized and presented on a unified platform~\cite{davidson2010cross}, improving content management efficiency and offering powerful support for cross-media creation and analysis. This flexibility not only expands \textit{ClueCart}'s application scope but also facilitates seamless transitions and in-depth interpretation across various storytelling formats.

\subsection{Design Implications}
\subsubsection{Comparative Analysis of Game Clue Collection Tools}
\par When comparing our mod with widely-used capture tools (P$1$, P$27$) like \textit{OBS}\footnote{\url{https://obsproject.com/}} or \textit{Fraps}\footnote{\url{https://fraps.com/}}, important design insights emerge. These tools, while popular for their simplicity, lack automatic in-game object detection and categorization, requiring creators to manually review and organize captured content. This can be particularly challenging for managing large volumes of assets ($34$ out of $40$). Our mod fills this gap by automatically detecting and classifying game objects during capture, significantly reducing manual effort and enhancing workflow efficiency. This suggests a need for integrating automated classification tools into future capture systems to streamline processes for creators. However, mod installation can be more complex than standalone tools, which highlights the importance of simplifying mod integration for less experienced users. Additionally, privacy concerns (P$14$, P$39$, P$40$) arise from automated data capture, making it crucial for developers to design privacy-first frameworks that secure user and game data. Looking ahead, integrating automated asset capture directly into the game's core design could enhance platform compatibility across systems like \textit{Nintendo Switch}\footnote{\url{https://www.nintendo.com/jp/hardware/switch/index.html}}, \textit{Xbox}\footnote{\url{https://www.xbox.com}}, and \textit{PlayStation}\footnote{\url{https://www.playstation.com}}, offering a seamless, secure experience while complying with platform regulations.

% \par \xiyuan{Each tool has its strengths and weaknesses, but our mod aims to address the limitations of current capture tools through future development. One potential solution involves collaborating with game developers to integrate asset capture and classification features directly into the game design process. This could ensure broader platform compatibility—including on closed systems like \textit{Nintendo Switch}, \textit{Xbox} and \textit{PlayStation}—and provide users with a more seamless, secure, and game-native experience. By integrating these features from the start, we can enhance creative workflows while safeguarding privacy and ensuring compliance with platform constraints.}

\subsubsection{Creator-Centric Clue Taxonomy}
\par By applying the proposed taxonomy to categorize game story clues, participants ($37$ out of $40$) reported significant improvements in their ability to comprehend and retrieve narrative elements. This taxonomy aligns closely with participants' intuitive workflows, indicating its potential for broader generalizability. For other developers and researchers, this approach provides a practical, adaptable framework that can be leveraged in diverse narrative-driven games. Its flexibility suggests that it could be extended beyond the contexts we studied, offering potential benefits for various story-rich games and even transmedia storytelling. While many games include built-in menu systems (e.g., in \textit{The Witcher 3: Wild Hunt}~\cite{gameWitcher3}),
% \footnote{\url{https://www.thewitcher.com/us/en/witcher3}})
these tend to focus on gameplay mechanics or achievements rather than supporting deeper narrative interpretation~\cite{kandelin2021creating}. By contrast, our taxonomy addresses a gap by providing creators with a dedicated tool for analyzing and organizing story elements, independent of the developer's original intent. This aligns with other work advocating for more story-centric design approaches in interactive media~\cite{kukka2017creator,saleh2019collaborative}.

% \par \xiyuan{By applying the proposed taxonomy to categorize collected game story clues, participants (Pxx, Pxx, ...) reported a marked improvement in their ability to understand and retrieve game clues. The taxonomy closely aligned with their intuitive workflows, suggesting it has a high degree of generalizability. This is particularly important as it demostrates not only its utility but also its flexibility in various narrative-driven tasks. While some games offer built-in, player-centric classification systems ([image examples]), these systems are often specific to the game's design and primarily reflect the perspectives of game developers rather than story interpretation creators. Such systems tend to focus on gameplay mechanics or achievements, leaving a gap in the organization of narrative elements for interpretive purposes. In contrast, our study's taxonomy fills this gap by providing a framework specifically tailored for creators who analyze and interpret stories. This taxonomy supports a more standardized, yet adaptable, method of categorizing story elements, enhancing both usability and accessibility across different game narratives.}

% % % % % % 写不动了% % % % % % 
% \subsubsection{Comparison between Miro and ClueCart (tbc)}
% % % % % % % % % % % % 

\subsubsection{Bridging Game Worlds with Real-World Context}
\par Integrating real-world contexts into game story interpretation has profound implications for content creation tools, particularly in their ability to support creators in producing deeper, more nuanced story interpretations~\cite{cesario2023exploring}. System designed to facilitate this process — such as LLM-powered contextualization tools — automatically surface symbolic meanings and cultural references, reducing the need for creators to have extensive prior knowledge in specific areas like history or mythology. For example, games like \textit{Dark Souls}~\cite{gameDarkSouls} are filled with complex references to medieval history, religious symbolism, and esoteric lore~\cite{smethurst2016mapping}. Traditionally, creators had to conduct deep research to interpret these clues (P$11$, P$15$, P$19$, P$37$). By leveraging LLMs, creators can access contextual explanations automatically, allowing them to focus on crafting richer, more multidimensional interpretations with out being hindered by the need for specialized knowledge. This design approach lowers barriers to content creation, enabling a wider range of creators to engage with and interpret complex game stories. By simplifying the process, these tools make game narratives more accessible and open to diverse interpretations, enhancing both creator workflows and the storytelling experience for audiences.

\subsubsection{\xiyuan{Balancing Structured Thinking and Creative Expression}}
\par \xiyuan{\textit{ClueCart} encourages creators to adopt a structured, systematic approach to story interpretation, promoting both creative and critical thinking. By linking clues and discovering connections between in-game clues and real-world concepts, creators can uncover narrative threads more comprehensively, enhancing creativity (P$15$, P$18$, P$33$, P$38$)~\cite{goldstein2001concept}. On the critical thinking side, \textit{ClueCart}'s tagging system and relationship visualization tools lead creators to assess how different clues contribute to the overarching story (P$13$, P$14$, P$16$, P$31$, P$39$). This approach helps spot gaps, inconsistencies, or overlooked connections, ultimately supporting a more coherent and logical interpretation of the narrative.}

\par \xiyuan{However, the concept map's structure introduces a trade-off between logical coherence and creative expression. While the tool supports clear, systematic thinking~\cite{gorman2015conceptual}, it limits the flexibility available in more freeform tools like \textit{Miro}, which allow creators to visually stretch, move, or reconfigure their ideas in more abstract, creative ways~\cite{white2012brainstorming}. In contrast, \textit{ClueCart}'s emphasis on organized, interconnected concepts, while valuable for ensuring logical clarity, may restrict the ability to visually explore the more organic, dynamic relationships that could emerge in less structured formats (P$5$, P$7$, P$21$). This trade-off highlights how a more structured, concept map approach, while effective in promoting logical analysis, can restrict the ability to visually represent creative, fluid interpretations. Future iterations could explore increasing flexibility in the concept map layout, enabling creators to blend the strengths of systematic thinking with greater room for creative expression.}

\subsection{Limitation and Future Work}
\par This work has several limitations. First, although we isolated three pivotal analytical perspectives—plot analysis, character relationships, and character portraits—creators did not express a strong preference for these categories in practice. Future work could explore more flexible, adaptive methods to narrative analysis. Second, the system's reliance on modding limits its accessibility on closed platforms like the \textit{Nintendo Switch}, \textit{Xbox} and \textit{PlayStation}. Future development should aim to integrate automatic clue collection directly into games for broader platform compatibility. Third, the manual exploration of clue relationships proved time-consuming for participants, suggesting future iterations could benefit from automated relationship detection to enhance efficiency. \xiyuan{Fourth, another limitation lies in the heavy reliance on LLM technology for tasks like keyword extraction and clue summarization. While these features streamline workflows, they may shift too much interpretive responsibility to the tool, potentially reducing user engagement. Future work could balance automation with user control by introducing customization options, such as adjustable extraction pro or customizable outputs, ensuring creators remain actively involved in the interpretive process.}

\par \xiyuan{Regarding the taxonomy, while it was designed to support narrative analysis, some participants found it slightly difficult to adapt to, indicating both a learning curve and limited flexibility across diverse game genres. Specially, the inability to add high-level categories further constrained users' ability to tailor the taxonomy to their needs, increasing the learning curve. Future work could focus on enhancing the taxonomy's intuitiveness and expanding its customization capabilities. By introducing a short tutorial to explain the taxonomy and integrating category customization functions into the modding, \textit{ClueCart} could better support flexible and adaptive narrative analysis while reducing user workload. This would allow users to create and manage additional high-level categories more efficiently, addressing the diverse needs of different game genres and improving the overall usability.}

\par Looking forward, \textit{ClueCart} could evolve into a dynamic, collaborative platform where creators share and organize clues in a wiki-like format. This would provide a resource pool for extensive game clues and allow creators to showcase their narrative interpretations. \textit{ClueCart} promotes creator-audience engagement by making interactions visible. As such, it fosters a more dynamic, involved community. \xiyuan{Additionally, future iterations could explore features supporting fan fiction creation, such as cross-genre character pairings and integration of assets from multiple game narrative universes, enabling innovative storytelling.} As the content grows, integrating machine learning could further personalize the experience by recommending clues based on each creator's unique style. This would enable innovative exploration of new concepts and enhance creative expression.

% add future work about extending taxonomy

\section{Conclusion}
\par This study introduced \textit{ClueCart}, a tool aimed at enhancing creativity by supporting game story interpretation through clue collection, classification, understanding, and inference. Insights gained from our workshop led to the creation of a creator-centric clue taxonomy, the development of an in-game mod for automatic clue collection, and the implementation of \textit{ClueCart} as a platform where creators can interpret game narratives. Results from the between-subjects study confirmed that \textit{ClueCart} improves the efficiency and quality of creators' story organization processes. It also validated the comprehensiveness and acceptance of the clue taxonomy. For future work, we plan to expand \textit{ClueCart}'s application, transforming it into an online platform that integrates multiple game story clues and reasoning resources, fostering the growth of the fan creation community.

\begin{acks}
\par We thank anonymous reviewers for their valuable feedback. This work is supported by grants from the National Natural Science Foundation of China (No. 62372298), Shanghai Engineering Research Center of Intelligent Vision and Imaging, Shanghai Frontiers Science Center of Human-centered Artificial Intelligence (ShangHAI), and MoE Key Laboratory of Intelligent Perception and Human-Machine Collaboration (KLIP-HuMaCo).
\end{acks}

\bibliographystyle{ACM-Reference-Format}
\bibliographystyle{ACM-Reference-Format}

%%%% bibliography starts here %%%%


\appendix
\newpage
\section{Prompt for Categorization, Description Generation, and Keyword Extraction}
\label{app:prompts}
\begin{lstlisting}
**Task:** Analyze screenshots from a game and categorize their content. For each category, perform specific analysis and generate structured output.
1. **Text-Only Screenshots**
   - **Task:** Extract all text visible in the screenshot.
   - **Output:** Summarize the main content in 30 words or fewer, focusing on keywords (up to three).
2. **NPC Dialogue Screenshots**
   - **Task:** Extract the full dialogue from NPC interactions.
   - **Output:** Summarize the conversation in 30 words or fewer and highlight up to three main keywords.
3. **Artifact-Related Screenshots**
   - Screenshots with text containing ``inventory'':
     - **Task:** Identify as an inventory page. Extract item names and descriptions.
     - **Output:** Provide a concise summary of the inventory content, emphasizing up to three main keywords.
   - Other artifact-related screenshots:
     - **Task:** Extract and list artifact names and descriptions.
     - **Output:** Summarize findings, with up to three main keywords.
4. **Environment Screenshots**
   - Applicable for screenshots without text or those containing the keywords: ``ENTER,'' ``CHALLENGE,'' or ``LISTEN.''
   - **Task:** Provide a one-sentence summary of the depicted environment.
   - **Output:** Include up to three main keywords relevant to the scene.

**General Guidelines:**
- Ignore any health indicators or character statuses shown in the top-left corner of screenshots.
- Summaries must not exceed 30 words per screenshot.

**Expected Output Format:**
- **Element:** Specify the type (e.g., ``text-only,'' ``dialogue,'' ``artifact,'' or ``environment'').
- **Description:** Include extracted text or summaries as per category.
- **Keywords:** List up to three keywords related to the content.
\end{lstlisting}
\newpage
\section{Prompt for Real-World Context}
\label{app:prompts4rec}
\begin{lstlisting}
Define the term ${keyword} as it relates to the game ${game}. Provide a concise explanation in 20 words or fewer.
\end{lstlisting}

\onecolumn
\section{Participants in the user study}
\label{app:user_study_participants}
\begin{table*}[ht]
		\caption{Detailed information of the participants in two user studies.}
		\label{tab:user_study_participants}
		\centering
  \resizebox{0.8\linewidth}{!}{
		\begin{tabular}{lllcccc}
			\toprule
			\textbf{Game} & \textbf{Tool} & \textbf{ID} & \textbf{Gender} & \textbf{Age} & \textbf{Platform} & \textbf{Frequency} \\
			\midrule
			        \multirow{20}{*}{Hollow Knight}  
        & \multirow{10}{*}{Miro} & P1  & Male   & 22  & Video Platforms                 & Seasonally     \\
        &                         & P2  & Male   & 23  & Video Platforms                 & Monthly         \\
        &                         & P3  & Female & 21  & Video Platforms \& Forums \& Game Wiki   & Seasonally    \\   
        &                         & P4  & Male   & 21  & Video Platforms                 & Monthly     \\
        &                         & P5  & Male   & 20  & Video Platforms                 & Weekly         \\
        &                         & P6  & Female & 21  & Video Platforms                 & Daily    \\   
        &                         & P7  & Male   & 23  & Video Platforms                 & Weekly     \\
        &                         & P8  & Male   & 22  & Video Platforms                 & Weekly         \\
        &                         & P9  & Female & 24  & Video Platforms \& Forums       & Weekly    \\   
        &                         & P10 & Female & 22  & Video Platforms \& Forums       & Weekly     \\
        \cmidrule{2-7}
        & \multirow{10}{*}{ClueCart} & P11 & Male   & 24  & Video Platforms \& Forums       & Monthly         \\
        &                         & P12 & Female & 20  & Video Platforms \& Forums       & Seasonally    \\   
        &                         & P13 & Female & 24  & Video Platforms                 & Weekly     \\
        &                         & P14 & Male   & 23  & Video Platforms \& Forums       & Monthly         \\
        &                         & P15 & Male   & 24  & Video Platforms                 & Seasonally    \\   
        &                         & P16 & Female & 22  & Video Platforms                 & Monthly     \\
        &                         & P17 & Male   & 22  & Video Platforms \& Forums       & Weekly         \\
        &                         & P18 & Male   & 20  & Video Platforms                 & Seasonally    \\   
        &                         & P19 & Female & 21  & Video Platforms \& Forums       & Monthly     \\
        &                         & P20 & Male   & 22  & Video Platforms \& Forums       & Monthly         \\ 	
        \midrule
			        \multirow{20}{*}{DNFTM}  
        & \multirow{10}{*}{Miro} & P21  & Male   & 28  & Video Platforms                 & Seasonally     \\
        &                         & P22  & Female & 21  & Forums                      & Weekly         \\
        &                         & P23  & Female & 22  & Video Platforms                 & Weekly    \\   
        &                         & P24  & Male   & 21  & Video Platforms \& Forums       & Weekly     \\
        &                         & P25  & Male   & 20  & Video Platforms                 & Weekly         \\
        &                         & P26  & Male   & 22  & Video Platforms \& Forums       & Monthly    \\   
        &                         & P27  & Male   & 22  & Video Platforms \& Forums       & Weekly     \\
        &                         & P28  & Female & 24  & Video Platforms \& Forums       & Monthly         \\
        &                         & P29  & Male & 24  & Video Platforms             & Weekly    \\   
        &                         & P30 & Female & 24  & Video Platforms                 & Monthly     \\
        \cmidrule{2-7}
        & \multirow{10}{*}{ClueCart} & P31 & Female & 24  & Video Platforms \& Forums       & Seasonally      \\
        &                         & P32 & Male   & 25  & Video Platforms \& Forums       & Monthly    \\   
        &                         & P33 & Female & 22  & Video Platforms                 & Seasonally     \\
        &                         & P34 & Male   & 21  & Video Platforms \& Forums       & Weekly         \\
        &                         & P35 & Female & 23  & Video Platforms                 & Seasonally    \\   
        &                         & P36 & Female & 23  & Video Platforms \& Forums       & Weekly      \\
        &                         & P37 & Male   & 21  & Video Platforms \& Forums       & Weekly         \\
        &                         & P38 & Female & 24  & Video Platforms                 & Monthly    \\   
        &                         & P39 & Female & 20  & Video Platforms                 & Seasonally     \\
        &                         & P40 & Male   & 29  & Video Platforms \& Forums       & Daily         \\  
        \bottomrule
		\end{tabular}
  }
\end{table*}

\newpage
\section{Evaluation Dimensions}
\label{app:dimension}

\begin{table*}[ht]
    \caption{Three adopted dimensions, along with their categories and corresponding aspects.}
    \centering
    \resizebox{1\linewidth}{!}{
    \begin{tabular}{p{3cm}lp{9.5cm}}
        \toprule
        \textbf{Dimensions} & \textbf{Categories} & \textbf{Aspect} \\
        \midrule
        \multirow{3}{=}{Perception} 
        & System Usability Scale & Gives a global view of subjective assessments of tool usability.  \\
        & \multirow{2}{*}{Creativity Support Index}  & Assesses how effectively the tool supports creative processes during task completion. \\
        \midrule
        \multirow{3}{=}{Features} 
        & Taxonomy & Facilitates \Classification{Classification \& Organization}, addressing C1 and C2 (\Classification{DR1}).\\[0.2em] \cmidrule(lr){2-3}
        & \multirow{2}{*}{Functionality}
            & Facilitates \Understanding{Clue Understanding}, addressing C3 and C4 (\Understanding{DR2}, \Understanding{DR3}). \\ [0.2em]
            & & Facilitates \Interpretation{Story Interpretation}, addressing C5 and C6 (\Interpretation{DR4}, \Interpretation{DR5}). \\ [0.1em]\cmidrule(lr){2-3}
        & Interaction & Assesses the overall system interaction design. \\
        \midrule
        \multirow{3}{=}{Story Interpretation\newline Assessment} 
        & Completion Time & Measures the time taken by participants to complete the task. \\
        & \multirow{2}{*}{Quality} 
            & Evaluates clarity, organization, and logical structure using a concept map evaluation checklist.  \\
        \bottomrule
    \end{tabular}}
    \label{tab:features}
\end{table*}

\section{System Usability and Creativity Support Results}
\label{app:perception}

\begin{figure*}[ht]
  \centering
  \includegraphics[width=\linewidth]
  {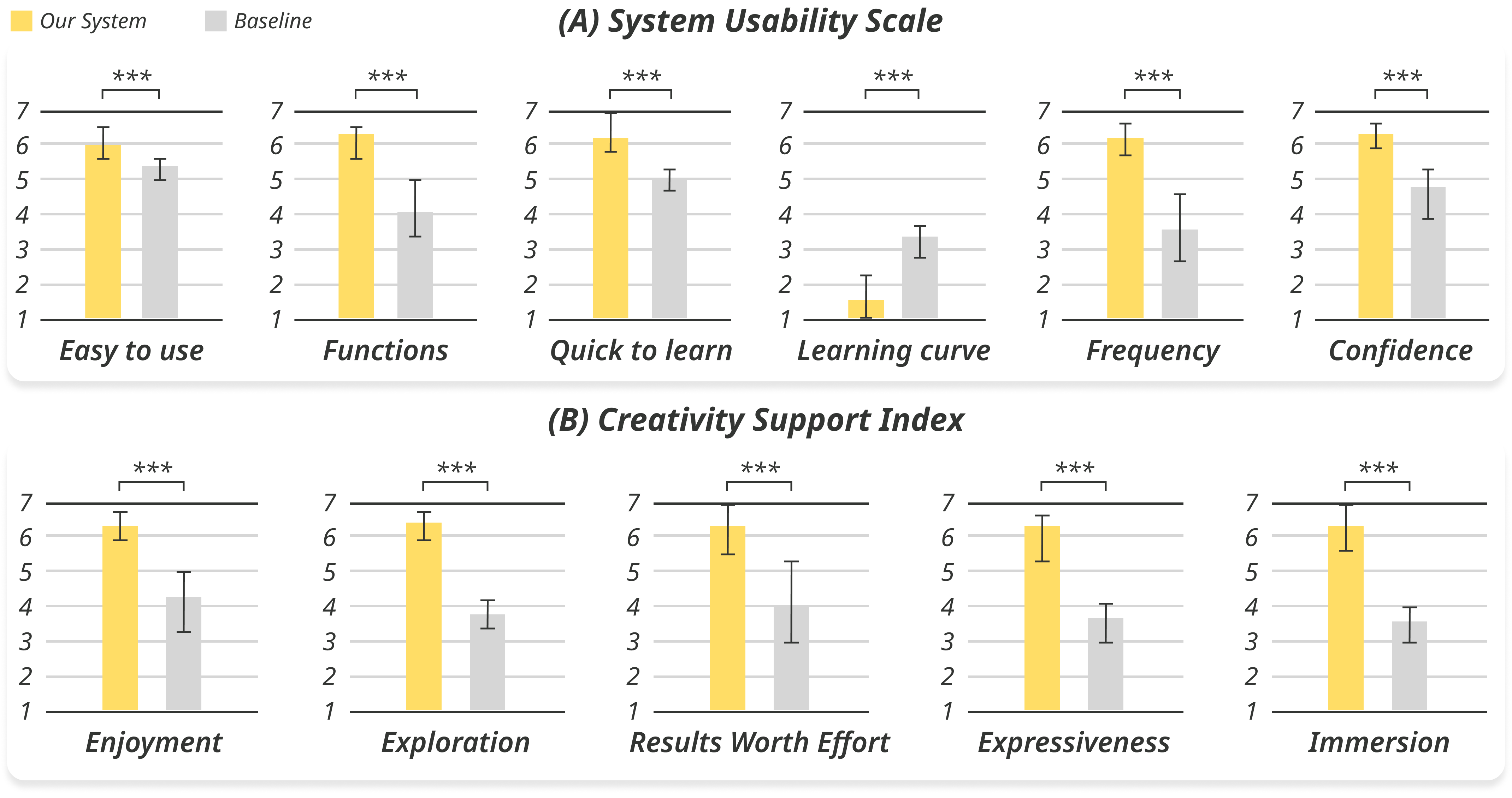}
  \caption[]{System Usability Scale and Creativity Support Index. \xiyuan{All the metrics show significant improvement compared with the baseline.}}
  \label{fig:sus_csi}
\end{figure*}

\newpage
\section{Questions about Perception of \textit{ClueCart}.}
\label{app:perception_questions}

\begin{table*}[ht]
  \caption{Questions about perception of \textit{ClueCart}.}
  \label{tab:perception_questions}
  \centering
  \resizebox{1\linewidth}{!}{
  \begin{tabular}{p{2cm}p{3cm}p{10cm}} % Adjust the width as needed
    \toprule
    \textbf{Categories} & \textbf{Factors} & \textbf{Questions}\\
    \midrule
    \multirow{6}{=}{System\newline Usability\newline Scale} 
      & Easy to use & I thought the system was easy to use.\\
      & Functions  & I found the various functions in this system were well integrated.\\
      & Quick to learn & I would imagine that most people would learn to use this system very quickly.\\
      & Learning curve & I needed to learn a lot of things before I could get going with this system.\\
      & Frequency & I think that I would like to use this system frequently.\\
      & Confidence & I felt very confident using the system.\\
    \midrule
    \multirow{5}{=}{Creativity\newline Support\newline Index} 
      & Enjoyment & I enjoyed using this system.\\
      & Exploration & The system helped me to track different ideas, outcomes, or possibilities.\\
      & Results Worth Effort & What I was able to produce was worth the effort I had to exert to produce it.\\
      & Expressiveness & The system allowed me to be very expressive.\\
      & Immersion & I became absorbed in the activity and forgot about the system that I was using.\\
    \bottomrule
  \end{tabular}}
\end{table*}

\section{Questions about Features of \textit{ClueCart}.}
\label{app:feature_questions}
\begin{table*}[ht]
  \caption{Questions about features of \textit{ClueCart}.}
  \label{tab:feature_questions}
  \centering
  \resizebox{1\linewidth}{!}{
  \begin{tabular}{p{2cm}p{4cm}p{10cm}} % Adjust the width as needed
    \toprule
    \textbf{Categories} & \textbf{Features} & \textbf{Questions}\\
    \midrule
    % \multirow{1}{*}{Taxonomy}
    %   & \Classification{DR1} Clue Categorization & \textit{ClueCart}'s clue categorization feature fits my usage habits.\\
    % \midrule
    % \multirow{5}{*}{Functionality}
    %   & \Understanding{DR2} Content Understanding & \textit{ClueCart} helps me quickly understand the content of each clue.\\
    %   & \Understanding{DR3} Relationship Understanding & \textit{ClueCart} helps me quickly understand the relationship between clues.\\
    %   & \Interpretation{DR4} Retrieving Clues & \textit{ClueCart} helps me quickly find the materials I need.\\
    %   & \Interpretation{DR5} New Insights from External Information & The external information search feature in \textit{ClueCart}'s Interpretation section helps me gain new insights.\\
    % \midrule
    \multirow{1}{*}{Taxonomy}
      & Clue Categorization & \textit{ClueCart}'s clue categorization feature fits my usage habits.\\
    \midrule
    \multirow{5}{*}{Functionality}
      & Content Understanding & \textit{ClueCart} helps me quickly understand the content of each clue.\\
      & Relationship Understanding & \textit{ClueCart} helps me quickly understand the relationship between clues.\\
      & Retrieving Clues & \textit{ClueCart} helps me quickly find the materials I need.\\
      & New Insights from External Information & The external information feature in \textit{ClueCart}'s Interpretation section helps me gain new insights.\\
    \midrule
    \multirow{1}{*}{Interaction}
      & Overall System Interaction & \textit{ClueCart}'s interactions help me organize clues and reason through the story.\\
    \bottomrule
\end{tabular}}
\end{table*}

\section{Questions about Story Interpretation Assessment of \textit{ClueCart}.}
\label{app:result_questions}

\begin{table*}[ht]
  \caption{Questions about story interpretation assessment of \textit{ClueCart}.}
  \label{tab:result_questions}
  \centering
  \resizebox{0.9\linewidth}{!}{
  \begin{tabular}{ll} % Adjust the width as needed
    \toprule
    \textbf{Factors} & \textbf{Questions}\\
    \midrule
      Clarity and Focus & The result is clear, legible, and focused.\\
       Accuracy and Organization & Information is clear, accurate, and well organized.\\
       Logical Structure & Concepts are linked logically and reflect the essential information about the story.\\
       Story Comprehension & Content is logically arranged to facilitate comprehension.\\
       Evidence of Topic & The result shows evidence of the topic.\\
    \bottomrule
  \end{tabular}}
\end{table*}









\end{document}